\documentclass[aps,prd,twocolumn,superscriptaddress,nofootinbib,amsmath,amsfonts
,preprintnumbers,notitlepage,10pt,english]{revtex4-1}
\pdfoutput=1
\usepackage{hyperref}
\usepackage{babel}
\usepackage{graphicx}
\usepackage{dcolumn}
\usepackage{bm}
\usepackage{cleveref}
\usepackage{color}
\usepackage[usenames,dvipsnames]{xcolor}
\usepackage[utf8]{inputenc}


\begin{document}

\title{eLISA eccentricity measurements as tracers of binary black hole formation}

\author{Atsushi Nishizawa}
\email{anishiza@olemiss.edu}
\affiliation{Department of Physics and Astronomy, The University of 
Mississippi, University, MS 38677, USA}
\author{Emanuele Berti}
\email{eberti@olemiss.edu}
\affiliation{Department of Physics and Astronomy, The University of 
Mississippi, University, MS 38677, USA}
\affiliation{CENTRA, Departamento de F\'isica, Instituto Superior
T\'ecnico, Universidade de Lisboa, Avenida Rovisco Pais 1,
1049 Lisboa, Portugal}
\author{Antoine Klein}
\email{aklein@olemiss.edu}
\affiliation{Department of Physics and Astronomy, The University of 
Mississippi, University, MS 38677, USA}
\affiliation{CENTRA, Departamento de F\'isica, Instituto Superior
T\'ecnico, Universidade de Lisboa, Avenida Rovisco Pais 1,
1049 Lisboa, Portugal}
\author{Alberto Sesana}
\email{asesana@star.sr.bham.ac.uk}
\affiliation{School of Physics and Astronomy, The University of Birmingham, 
Edgbaston, Birmingham B15 2TT, UK}

\pacs{04.30.Tv,04.25.Nx,97.60.Lf}

\date{\today}

\begin{abstract}
  Up to hundreds of black hole binaries individually resolvable by
  eLISA will coalesce in the Advanced LIGO/Virgo band within ten
  years, allowing for multi-band gravitational wave observations.
  Binaries formed via dynamical interactions in dense star clusters
  are expected to have eccentricities $e_0\sim 10^{-3}$--$10^{-1}$ at
  the frequencies $f_0=10^{-2}$~Hz where eLISA is most sensitive,
  while binaries formed in the field should have negligible
  eccentricity in both frequency bands. We estimate that eLISA should
  always be able to detect a nonzero $e_0$ whenever
  $e_0\gtrsim 10^{-2}$; if $e_0\sim 10^{-3}$, eLISA should detect
  nonzero eccentricity for a fraction $\sim 90\%$ ($\sim 25\%$) of
  binaries when the observation time is $T_{\rm obs}=5$ ($2$) years,
  respectively. Therefore eLISA observations of black hole binaries
  have the potential to distinguish between field and cluster
  formation scenarios.
\end{abstract}

\maketitle

\section{Introduction}

With the detection of gravitational waves (GWs) by the LIGO/Virgo
scientific collaboration~\cite{Abbott:2016blz}, black hole (BH)
binaries have entered the realm of observational astronomy.  The first
detected binary system (GW150914) has source-frame component masses
$(m_1, m_2) = (36^{+5}_{-4}, 29^{+4}_{-4})~M_\odot$, resulting in a
merger remnant of mass $62^{+4}_{-4}~M_\odot$.
Its estimated luminosity distance is
$D_{\rm L}=410^{+160}_{-180}$~Mpc, corresponding to a redshift
$z=0.09^{+0.03}_{-0.04}$~\cite{TheLIGOScientific:2016wfe}.
The trigger LVT151012 is also likely to be a binary BH system with
masses $(m_1, m_2) = (23^{+18}_{-5}, 13^{+4}_{-3})M_\odot$ and
luminosity distance $D_{\rm L}=1.1^{+0.5}_{-0.5}$~Gpc.
These early GW observations set lower bounds on binary BH merger
rates~\cite{Abbott:2016nhf}, raising interesting questions on the
formation mechanism of compact binary systems. As summarized in the
LIGO/Virgo collaboration paper discussing the astrophysical
implications of the discovery~\cite{TheLIGOScientific:2016htt}, BH
binary mergers similar to GW150914 can either result from the
evolution of isolated binaries in galactic fields or from dynamical
interactions in young and old dense star clusters (see
\cite{Postnov:2014tza,Benacquista:2011kv} for reviews of these
formation scenarios).
 
Sesana ~\cite{Sesana:2016ljz} showed that up to hundreds of
GW150914-like BH binaries individually resolvable by a space-based
detector such as eLISA~\cite{AmaroSeoane:2012km} will coalesce in the
LIGO band within ten years.  eLISA observations can identify the time
and location of the merger with uncertainties in the merger time
smaller than $\sim 10$~s, and sky localization accuracies that in many
cases are better than $1$~deg$^2$. This will allow
  multi-wavelength electromagnetic telescopes to point the GW event in
  advance and to constrain models of electromagnetic emission
  associated with BH binary mergers. Furthermore, BH binaries that
span both the eLISA and Advanced LIGO frequency bands can yield
stringent tests of modified theories of gravity that predict
propagation properties different from general
relativity~\cite{TheLIGOScientific:2016src,Yunes:2016jcc}, and in
particular of theories allowing for dipolar radiation in BH
binaries~\cite{Barausse:2016eii}.

The GW150914 signal does not set strong bounds on the eccentricity $e$
of the binary. Ref.~\cite{TheLIGOScientific:2016wfe} quotes a
preliminary constraint of $e<0.1$ at $f=10$~Hz.  It is unlikely that
Advanced LIGO observations may use eccentricity measurements to
differentiate between the field and cluster scenarios: as shown
e.g. in Fig.~3 of Ref.~\cite{TheLIGOScientific:2016htt}, binaries in
the LIGO band will almost always be circular. Earth-based GW
observations could only differentiate between field and cluster
formation by looking at spin dynamics (see
e.g.~\cite{Gerosa:2013laa}), redshift distribution and possibly kicks.

However binaries formed in clusters -- unlike binaries formed in the
field -- should have non-negligible eccentricity in the eLISA
band. Here we show that eLISA could measure the eccentricity of BH
binaries in the last few years or months of their inspiral,
constraining their formation mechanism.  As a byproduct, we also show
how eccentricity affects the estimation of other binary parameters
(masses, merger time, distance and sky location).

The possibility of multi-band detections of eccentric
intermediate-mass BH binaries by Earth- and space-based detectors was
pointed out in a series of papers by Amaro-Seoane et
al.~\cite{AmaroSeoane:2009yr,AmaroSeoane:2009cg,AmaroSeoane:2009ui},
but those papers focused on BH binaries with much larger total mass.
Seto~\cite{Seto:2016wom} recently studied eccentric BH binaries of the
GW150914 type in the eLISA band, but the focus of his work was
considerably different from ours. He considered monochromatic sources
at frequencies $\sim 0.1-1$~mHz, which have negligible frequency
evolution, and for which the merger will not be visible in the
Advanced LIGO band. On the contrary we focus on binaries that evolve
rapidly in the high-frequency band of the eLISA sensitivity window,
possibly merging in the Advanced LIGO band. Seto used the quadrupole
formula to estimate the signal (which for $e\lesssim 0.1$ is dominated
by the second harmonic, i.e. by GWs emitted at twice the orbital
frequency) and estimated the binary eccentricity from the
characteristic amplitude of the third harmonic of the signal. We use
general relativistic waveform models and a Fisher matrix analysis to
estimate errors in the measurement of the eccentricity and of other
parameters characterizing the source (masses, merger time, distance
and sky location). We work within the small-eccentricity waveform
generation formalism proposed in~\cite{Yunes:2009yz} and further
developed in~\cite{Tanay:2016zog}, which is adequate to address the
present problem, but we note that various groups have recently made
progress in the development of models for the generation, detection
and parameter estimation of GWs from eccentric binaries (see
e.g.~\cite{Martel:1999tm,Mikoczi:2012qy,Huerta:2014eca,Sun:2015bva,Forseth:2015oua,Hopper:2015icj,Moore:2016qxz,Loutrel:2016cdw}).

In the rest of this introduction we review some literature on BH
formation channels and merger rates, including recent papers that were
not included in the LIGO review on this topic~\cite{Abadie:2010cf}, to
justify our statement that field binaries should typically be
circular, while binaries formed in clusters may have residual
eccentricities.  A more realistic study would require astrophysical
models of the mass, spin and eccentricity distribution of BH binaries
in both formation channels and Bayesian model
selection~\cite{Stevenson:2015bqa}; such an analysis is beyond the
scope of this paper, where we focus mostly on the preliminary issue of
parameter estimation accuracy. Then we present an executive summary of
our main results on eLISA measurements of eccentricity. Finally we
outline the plan of the paper for the reader's convenience.

\subsection{Black hole formation channels}

\noindent {\bf \em Field binaries.}  Tutukov and Yungelson studied the
evolution of isolated massive binaries before the discovery of the
binary pulsar and predicted the formation of merging binary compact
objects composed of neutron stars (NSs) and/or
BHs~\cite{Tutukov:1973,Tutukov:1993}. Some early population studies
even predicted that binary BH mergers could dominate detection rates
for ground-based GW detectors~\cite{Lipunov:1997gh}. Several groups
made predictions on the relative rates of BH-BH, BH-NS and NS-NS
binaries over the
years~\cite{Bethe:1998bn,Grishchuk:2000gh,Nelemans:2001hp,Belczynski:2001uc,Voss:2003ep,DeDonder:2004cx,Belczynski:2005mr,O'Shaughnessy:2006wh,Kalogera:2006uj}. All
of these predictions were largely uncertain, but as late as 2014 some
studies concluded that BH-BH binary detection rates would be
negligible for Advanced LIGO~\cite{Mennekens:2014}.

Belczynski et al.~\cite{Belczynski:2010tb} pointed out that BH-BH
binaries could dominate Advanced LIGO detection rates if a significant
fraction of stars form in low-metallicity environments.  This claim
was refined in subsequent work using the {\sc Startrack} code with
various prescriptions for common envelope evolution, BH kicks and
gravitational
waveforms~\cite{Dominik:2012kk,Dominik:2013tma,Belczynski:2014iua,Dominik:2014yma,Belczynski:2015tba,Belczynski:2016obo},
as well as various prescriptions for metallicity evolution as a
function of redshift. These works consistently predicted that BH
mergers should dominate the rates, and that large-mass BH binaries
(including total masses $\sim 60M_\odot$ and above) should be
detectable in large numbers out to $z\sim 2$.  Notably, before the
detection of GW150914 Belczynski et al.~\cite{Belczynski:2015tba}
found that ``the most likely sources to be detected with the advanced
detectors are massive BH-BH mergers with total redshifted mass
$\sim 30-70 M_\odot$.''

Similar conclusions were reached using other population synthesis
codes~\cite{Spera:2015,Eldridge:2016ymr}.  Eldridge and
Stanway~\cite{Eldridge:2016ymr} found that GW150914 has a low
probability of arising from a stellar population with initial
metallicity $Z \gtrsim 0.01$ (or $Z \gtrsim 0.5Z_\odot$); when
$Z=10^{-4}$, a large fraction ($\sim 26\%$) of binary BH mergers is
expected to have masses compatible with the GW measurement.
Other groups suggested that common envelope evolution may not be the
only way to form massive BHs. Another channel involves massive, tight
binaries where mixing induced by rotation and tides transports the
products of hydrogen burning throughout the stellar envelopes,
enriching the entire star with helium and preventing the build-up of
an internal chemical
gradient~\cite{Mandel:2015qlu,deMink:2016vkw,Marchant:2016wow}. In
these scenarios there would never be a giant phase: both stars would
stay within their Roche lobes and eventually form massive BHs, because
the cores that collapse would be large.
Yet another scenario invokes a Population III origin for massive BH
binaries~\cite{Kinugawa:2014zha,Kinugawa:2015nla,Inayoshi:2016hco},
but semi-analytical models suggest that the probability of GW150914
having formed in the early Universe is
$\sim 1\%$~\cite{Hartwig:2016nde}.

The key point for us is that {\em BH binaries produced in the field
  are expected to be circular in both the Advanced LIGO and eLISA
  bands}. Typical eccentricity distributions for BH binaries at
frequencies $\sim 0.3$~Hz are shown in Fig.~5
of~\cite{Kowalska:2010qg}; predicted values are in the range
$10^{-6}\lesssim e \lesssim 10^{-4}$. Massive BH binaries of interest
for multi-band astronomy are at the heavy end of the mass spectrum, so
they should receive small kicks (see e.g. Sec.~6
of~\cite{Belczynski:2015tba}) and be on the small-eccentricity side of
the distributions predicted in~\cite{Kowalska:2010qg}.  For all
practical purposes, massive BH binaries formed in the field can be
assumed to be circular in the eLISA band.

\noindent
{\bf \em Dense star clusters.}
A different scenario for binary BH formation involves dense star
clusters~\cite{Kulkarni:1993fr,Sigurdsson:1993zrm}. In these
environments BHs quickly become the most massive objects. They sink
towards the cluster core, form pairs through dynamical interactions,
and they are most commonly ejected in binary configurations with
inspiral times shorter than the age of the Universe. This basic
scenario was refined by various
authors~\cite{PortegiesZwart:1999nm,Gurkan:2005xz,Gultekin:2006,Fregeau:2006yz,
Miller:2008yw,O'Leary:2008xt,Moody:2008ht,Downing:2010,Downing:2011,
Morscher:2012se,Goswami:2013aha,Ziosi:2014sra,Rodriguez:2015oxa,
Rodriguez:2016kxx,O'Leary:2016qkt,Bartos:2016dgn,Chatterjee:2016hxc}.

A dynamical effect that can produce large eccentricities in the LIGO
band is the Kozai mechanism~\cite{Wen:2002km}. Recent studies of
Kozai-Lidov resonances showed that binary BH mergers may be more
likely inside the radius of influence of supermassive BHs in galactic
centers~\cite{Antonini:2012ad,VanLandingham:2016ccd} or in
hierarchical triples~\cite{Antonini:2015zsa,Samsing:2013kua},
More work is required to understand whether these events can lead to
rates comparable to the other formation channels, and also to
establish the conditions (masses, inclinations, semi-major axes and
eccentricities of both the inner and outer binary) that could result
in non-negligible eccentricities in the eLISA band.

Some predictions for the eccentricity distribution of dynamically
formed binaries can be found in Fig.~10
of~\cite{Rodriguez:2016kxx}. The eccentricity at $10~$Hz of BH
binaries merging at $z<1$ in the capture scenario peaks at
$e=10^{-6}$, with most of the sources having $e<10^{-5}$. The classic
results by Peters and Mathews~\cite{Peters:1963ux} imply that, so long
as $e\ll 1$, $e\sim f^{-19/18}\approx f^{-1}$ (see e.g. Fig.~1 of
\cite{Enoki:2006kj}).  Here we focus on sources emitting at
$f>f_0=10^{-2}$~Hz in the eLISA band. Their typical eccentricity at
frequency $f\sim f_0$ is thus $e\sim 10^{-3}$, with most sources
having $e\lesssim 10^{-2}$. Almost all relevant eLISA sources (both
resolvable and unresolvable) are at $f>10^{-3}$~Hz, and their expected
eccentricity is $e\lesssim 0.1$.  These numbers are large enough to
require eccentric templates for matched filtering, but the amplitude
and phasing of the signal for binaries with $e\lesssim 0.1$ can be
treated in a small-eccentricity approximation.
To summarize: extrapolating the results in
Ref.~\cite{Rodriguez:2016kxx} to lower frequencies, we expect
dynamically formed BH binaries to have small but non-negligible
eccentricities $e \lesssim 0.1$ in the eLISA band, and therefore a
small-eccentricity approximation is adequate to study this problem.

\subsection{Executive summary}

Consider a binary system with component masses (in the source frame)
$m_1$ and $m_2$, total mass $M=m_1+m_2$, symmetric mass ratio
$\eta=m_1 m_2/M^2$ and chirp mass ${\cal M}=\eta^{3/5} M$.
Assume that the binary is located at redshift $z$ -- or equivalently,
for a given cosmological model, at luminosity distance $D_L=D_L(z)$ --
so that the redshifted chirp mass ${\cal M}_z=(1+z) {\cal M}$, the
redshifted total mass $M_z=(1+z) M$, and similarly for the other mass
parameters. 
Two angles $(\bar{\theta}_{\rm S},\bar{\phi}_{\rm S})$ specify the
direction of the source in the solar barycenter frame, and for
convenience we introduce $R=1\,{\rm AU}$. Let $t_c$ be the coalescence
time, $\phi_c$ the coalescence phase, $\mathbf{L}$ the binary's
orbital angular momentum vector (with
$\hat{\mathbf{L}}=\mathbf{L}/|\mathbf{L}|$ the corresponding unit
vector), and $\hat{\mathbf{N}}$ a unit vector pointing in the source
direction as measured in the solar barycenter frame.
Furthermore, let $\chi=f/f_0$ be the frequency normalized to a
reference frequency -- here chosen to be $f_0=10^{-2}\,{\rm Hz}$ --
where the eccentricity is $e(f_0)=e_0$, and introduce the standard
post-Newtonian (PN) parameter $x=(\pi M_z f)^{2/3}$.

We model eLISA as two independent interferometers with non-orthogonal
arms. The sky-averaged noise power spectral density for each of the
two interferometers is denoted by N$i$A$j$, as
in~\cite{Klein:2015hvg}; here $i=1,\,2$ refers to different
acceleration noise baselines, and $j=1,\,5$ denotes different
armlengths (1 or 5~Gm). The observation time $T_{\rm obs}$ is chosen
to be either 5 or 2 years. This choice significantly affects the
signal-to-noise ratio (SNR): if, following~\cite{Sesana:2016ljz}, we
adopt a fiducial 5-year observation time and assume that the binary
merges at the end of the observation, the initial frequency of the
binary will be
\begin{equation}
f_{\rm{min}} = 0.015 \left( \frac{30\,M_{\odot}}{{\cal M}_z} \right)^{5/8} 
\left( \frac{5\,{\rm{yr}}}{T_{\rm obs}} \right)^{3/8} \; {\rm{Hz}}\;,
\label{eq1}
\end{equation}
where we scaled the result by the estimated redshifted chirp mass of
GW150914. 
Our SNR and Fisher matrix calculations are truncated at a maximum
frequency $f_{\rm max}=1\,{\rm Hz}$, beyond which the eLISA noise is
not expected to be under control.
 
\begin{figure}[t]
\begin{center}
\includegraphics[width=\columnwidth]{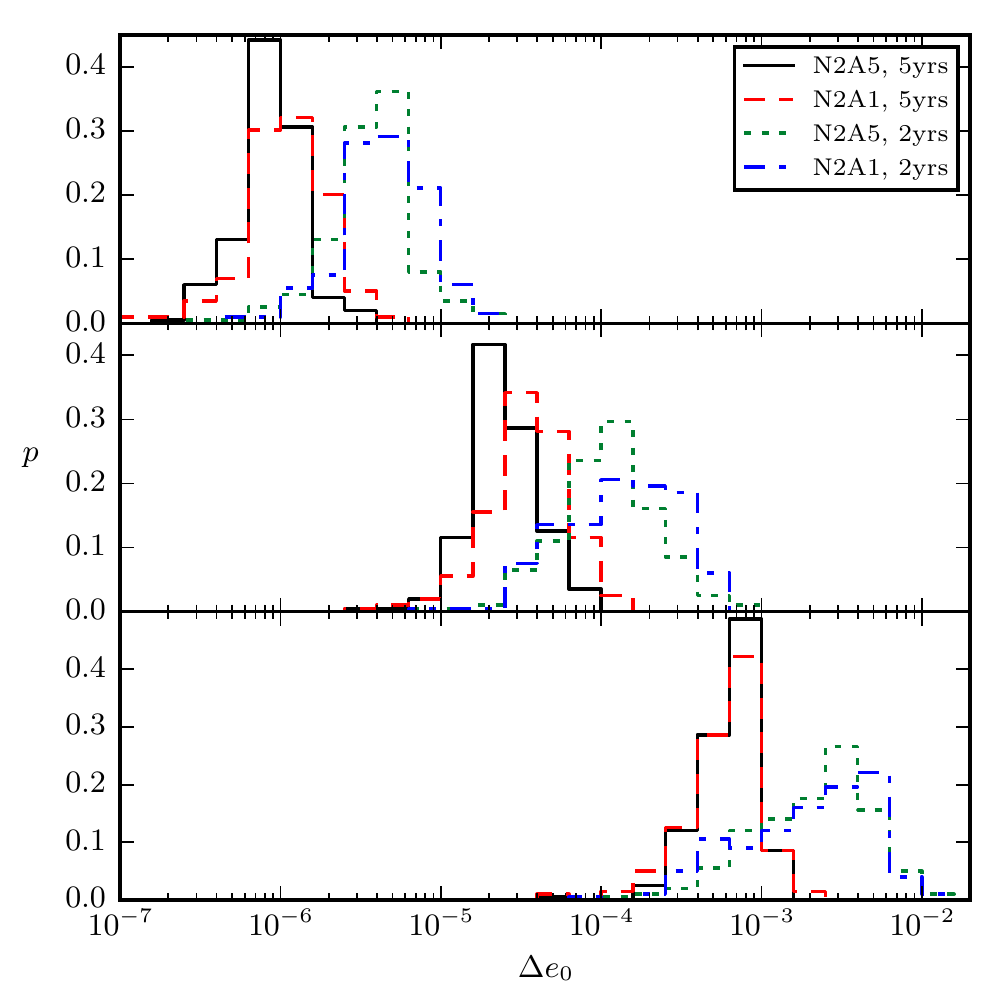}
\caption{Parameter estimation errors on the eccentricity $e_0$ at
  frequency $f_0=10^{-2}$~Hz using ``full eccentric'' waveforms for
  nonspinning binaries. Different panels refer to catalogs with
  $e_0 = 0.1$, $0.01$ and $0.001$ (from top to bottom). The various
  linestyles refer to different noise curves and observation times:
  N2A5 and $T_{\rm obs}=5\,{\rm yrs}$ (solid black), N2A1 and
  $T_{\rm obs}=5\,{\rm yrs}$ (dashed red), N2A5 and
  $T_{\rm obs}=2\,{\rm yrs}$ (dotted green), N2A1 and
  $T_{\rm obs}=2\,{\rm yrs}$ (dash-dotted blue).}
\label{fig:de0}
\end{center}
\end{figure} 

\begin{figure}[t]
\begin{center}
\includegraphics[width=0.9\columnwidth]{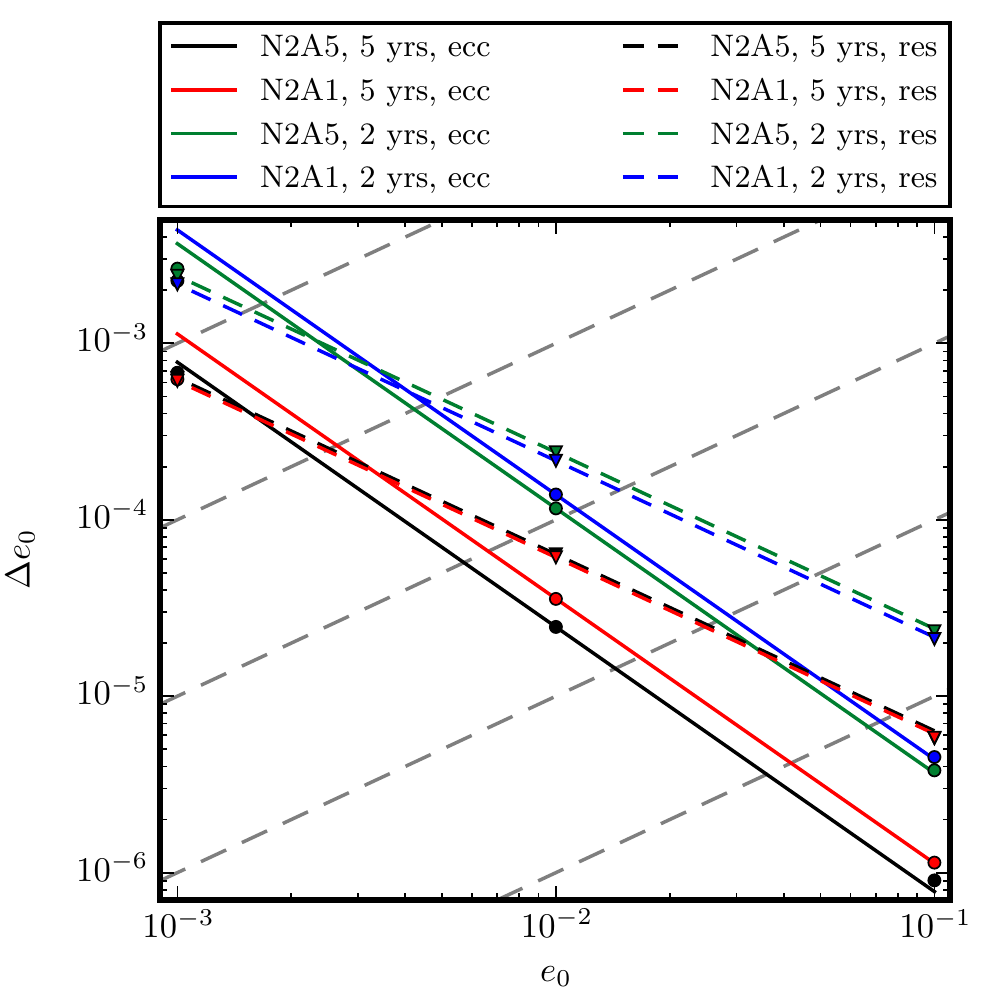}
\caption{Scaling of the median error $\Delta e_0$ with $e_0$ for the
  N2A5 noise model. Dashed (solid) lines correspond to restricted
  (full) eccentric waveforms, respectively. Colors refer to N2A5 and
  $T_{\rm obs}=5\,{\rm yrs}$ (black), N2A1 and
  $T_{\rm obs}=5\,{\rm yrs}$ (red), N2A5 and
  $T_{\rm obs}=2\,{\rm yrs}$ (green), N2A1 and
  $T_{\rm obs}=2\,{\rm yrs}$ (blue). From top left to bottom right,
  gray long-dashed lines correspond to $\Delta e_0/e_0=1$, $10^{-1}$,
  $10^{-2}$, $10^{-3}$, $10^{-4}$.}
\label{fig:scalings2}
\end{center}
\end{figure} 
 
Our main results on eccentricity measurements are summarized in
Figs.~\ref{fig:de0} and \ref{fig:scalings2}. Their behavior can be
understood, at least qualitatively, using simple scaling
arguments. Neglecting correlations between parameters, in a Fisher
matrix approximation the error on $e_0$ is
\begin{equation}
\Delta e_0 \sim \left[ f \frac{|\partial_{e_0} \tilde{h}|^2}{S_h} 
\right]^{-1/2} \;,
\label{eq11}
\end{equation}
where $\tilde{h}$ denotes the Fourier transform of the GW amplitude
and $S_h(f)$ is the noise power spectral density of the detector.  To
leading order in a small-eccentricity expansion (what we call the
``restricted eccentric waveform'' in Section~\ref{sec:resecc} below)
and in the stationary phase approximation, corrections due to the
eccentricity enter only in the GW phase through the term proportional
to $e_0^2$ in Eq.~(\ref{psi:resecc}) below, and therefore
$\partial_{e_0} \tilde{h} = {\cal M}_z^{-5/6} f^{-89/18} e_0$.
Let us approximate the frequency dependence of the noise power
spectral density by a power law, $S_h\sim f^{2\alpha}$. Since the
dominant contribution to the Fisher matrix comes from the lowest
frequencies, from Eq.~(\ref{eq11}) we have
\begin{align}
\Delta e_0 &\sim {\cal M}_z^{5/6} f_{\rm min}^{40/9+\alpha} e_0^{-1}
  \nonumber \\
&\sim {\cal M}_z^{-5(28+9\alpha)/72} T_{\rm obs}^{-(40+9\alpha)/24} e_0^{-1}\;,
\end{align}
where on the second line we estimated $f_{\rm min}$ for a given
observation time $T_{\rm obs}$ using the quadrupole formula
\eqref{eq1} for a circular binary.
In summary, to leading order we expect a rough scaling law of the form
\begin{equation}
\label{scalinge0an}
\Delta e_0 \sim {\cal M}_z^{-\gamma_m} T_{\rm obs}^{-\gamma_t} e_0^{-\gamma_e} \;,
\end{equation}
with $(\gamma_m,\gamma_t)=(2.57,2.04)$ for $\alpha=1$ (N2A5 and N2A1,
2yrs), $(\gamma_m,\gamma_t)=(2.19,1.82)$ for $\alpha=0.4$ (N2A1,
5yrs) 
and $\gamma_e=1$. Note that $\alpha$ depends not only on the
noise curve, but also on $f_{\rm min}$, that is lower for longer
$T_{\rm obs}$: the frequency dependence of the eLISA noise curve is
flatter when we consider N2A1 and a 5-year observation time.
 
This rough approximation will break down when the SNR is small (so the
Fisher matrix approximation is invalid), correlations cannot be
neglected (as is the case for the ``restricted'' eccentric waveform), or
eccentricities are too small and therefore not measurable.
In practice we carry out numerical calculations using the ``full''
eccentric waveform described in Section~\ref{sec:NLOecc}
below. Obtaining analytical estimates in this case is more complicated
due to the existence of frequency sidebands, but by fitting our
numerical data we found that the scaling law with ${\cal M}_z$ holds
well also for these full eccentric waveforms. Because of the breaking
of some parameter degeneracies, the scaling with $e_0$ is modified
from the previous simple prediction: $\gamma_e \approx 1.5$ for
$e_0 > 0.01$.
A more accurate scaling law obtained by fitting our numerical data is
\begin{align}
\label{scalinge0num}
\Delta e_0 \approx \epsilon_0 \left( \frac{d_L}{400\,{\rm Mpc}} \right) \left(  \frac{30\,M_{\odot}}{{\cal M}_z} \right)^{\gamma_m} \left(  \frac{0.1}{e_0} \right)^{\gamma_e} \;,
\end{align}
where the fitting parameters ($\epsilon_0,\,\gamma_m,\,\gamma_e$) are
listed in Table~\ref{tab:fite}. This scaling is further illustrated in
Fig.~\ref{fig:scalings2}.

The simple scalings of Eqs.~\eqref{scalinge0an} and
\eqref{scalinge0num} are helpful to understand the numerical results
shown in Fig.~\ref{fig:de0}. The error $\Delta e_0$ gets larger with
decreasing eccentricity: when $e_0\sim 0.1$ the typical error is
$\Delta e_0\approx 10^{-6}$, but when when $e_0\sim 0.001$ the typical
error $\Delta e_0\sim e_0$. For a given noise curve (N2A5 or N2A1), as
expected, longer observation times lead to smaller errors. The effect
of changing the armlength is sensibily milder, but (everything else
being equal) 5~Gm configurations (A5) yield slightly smaller errors
than 1~Gm configurations (A1).

Recall from our previous discussion that binaries formed in dense star
clusters are expected to have eccentricities
$10^{-3}\lesssim e_0\lesssim 10^{-2}$ at the frequencies
$f_0=10^{-2}$~Hz where eLISA is most sensitive, while binaries formed
in the field should have negligible eccentricity
$10^{-6}\lesssim e \lesssim 10^{-4}$ at these frequencies. eLISA
should always be able to detect a nonzero $e_0$ whenever
$e_0\gtrsim 10^{-2}$; if $e_0\sim 10^{-3}$, we find that eLISA will
detect nonzero eccentricity for a fraction $\sim 90\%$ ($\sim 25\%$)
of binaries when $T_{\rm obs}=5$ ($2$) years, respectively. Therefore
eLISA observations of GW150914-like BH binaries have the potential to
distinguish between field and cluster formation scenarios. This is the
main result of our paper.

\begin{table}[t]
\begin{center}
\begin{tabular}{ccccc}
noise &$T_{\rm obs}$ & \; $\epsilon_0$ \; & \; $\gamma_m$ \; & \;$\gamma_e$ \; \\
\hline  
\hline 
N2A1 & 2yr & $1.0 \times 10^{-5}$ & 2.57 & 1.5 \\
N2A1 & 5yr & $2.2 \times 10^{-6}$ & 2.19 & 1.5 \\
N2A5 & 2yr & $6.5 \times 10^{-6}$ & 2.57 & 1.5 \\
N2A5 & 5yr & $9.0 \times 10^{-7}$ & 2.57 & 1.5 \\
\hline 
\hline 
\end{tabular}
\end{center}
\caption{Fitting parameters in the scaling relation of 
Eq.~\eqref{scalinge0num}.}
\label{tab:fite}
\end{table}

\begin{figure*}[t]
\begin{center}
\includegraphics[width=0.32\textwidth]{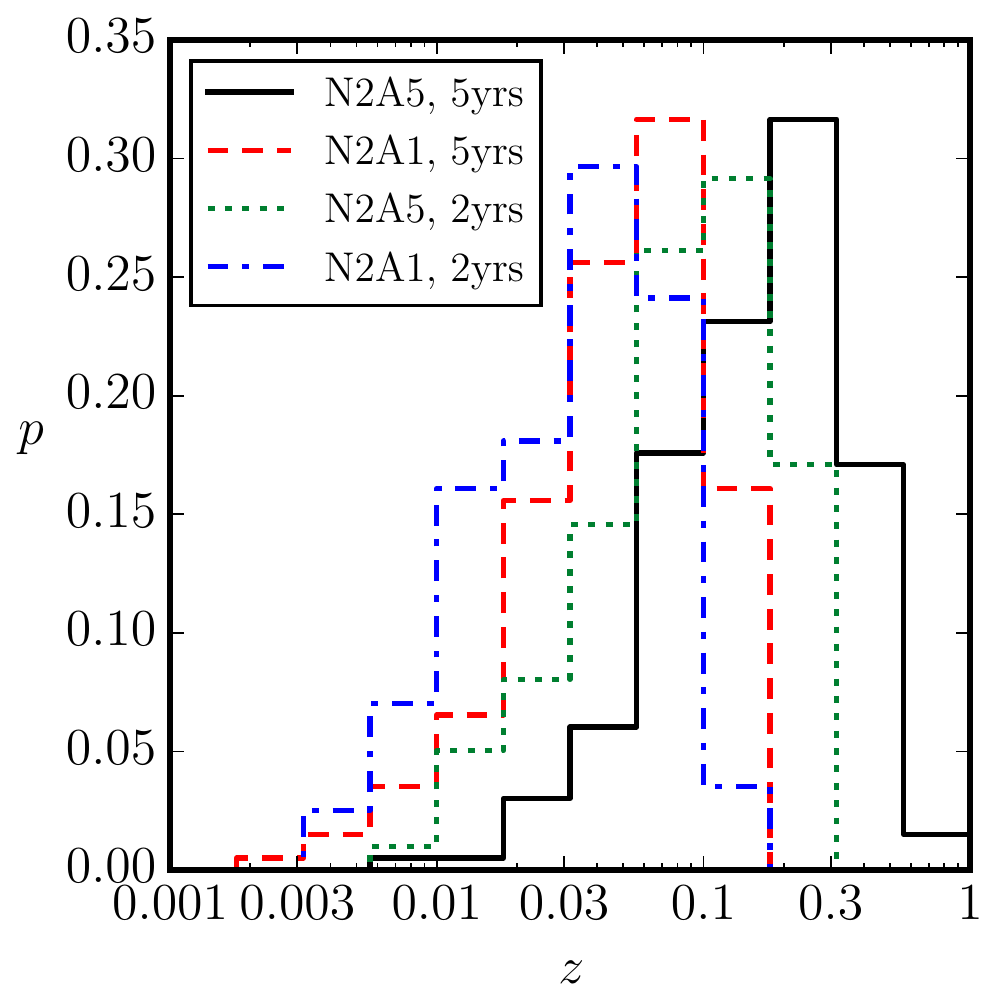}
\includegraphics[width=0.32\textwidth]{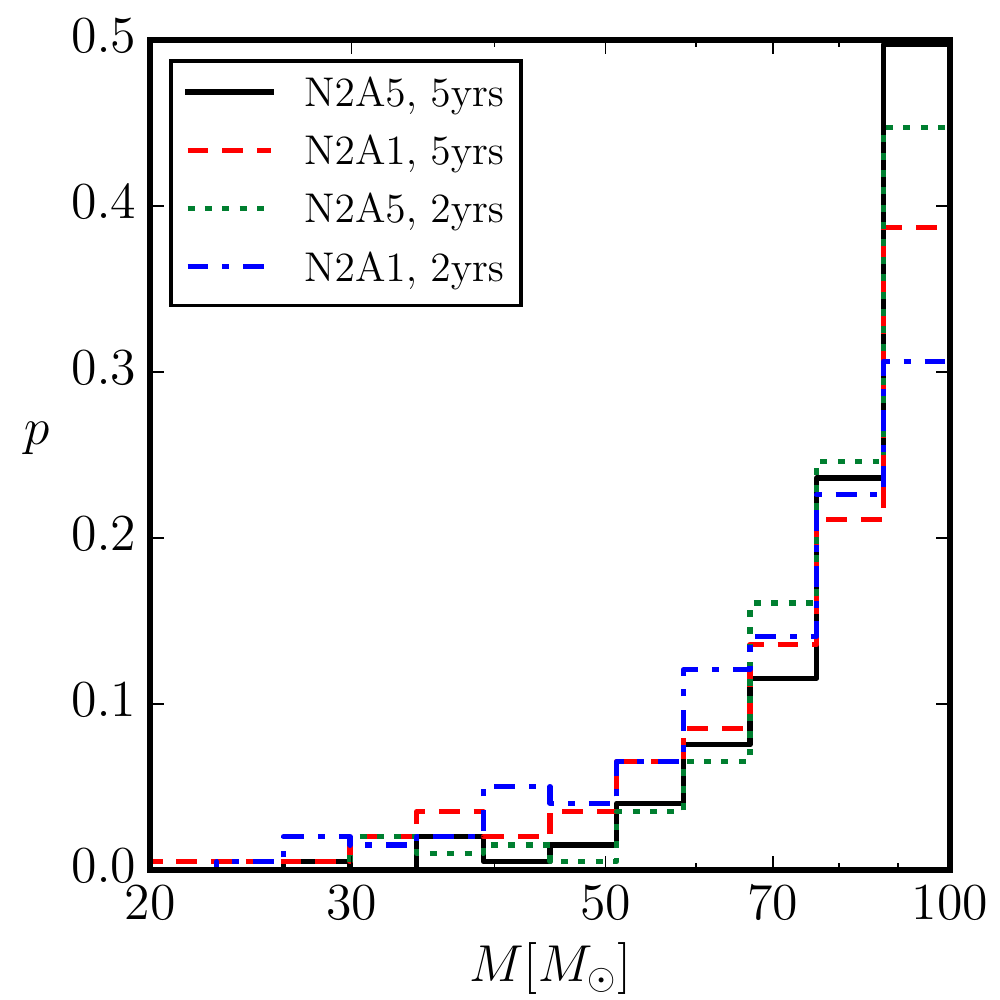}
\includegraphics[width=0.32\textwidth]{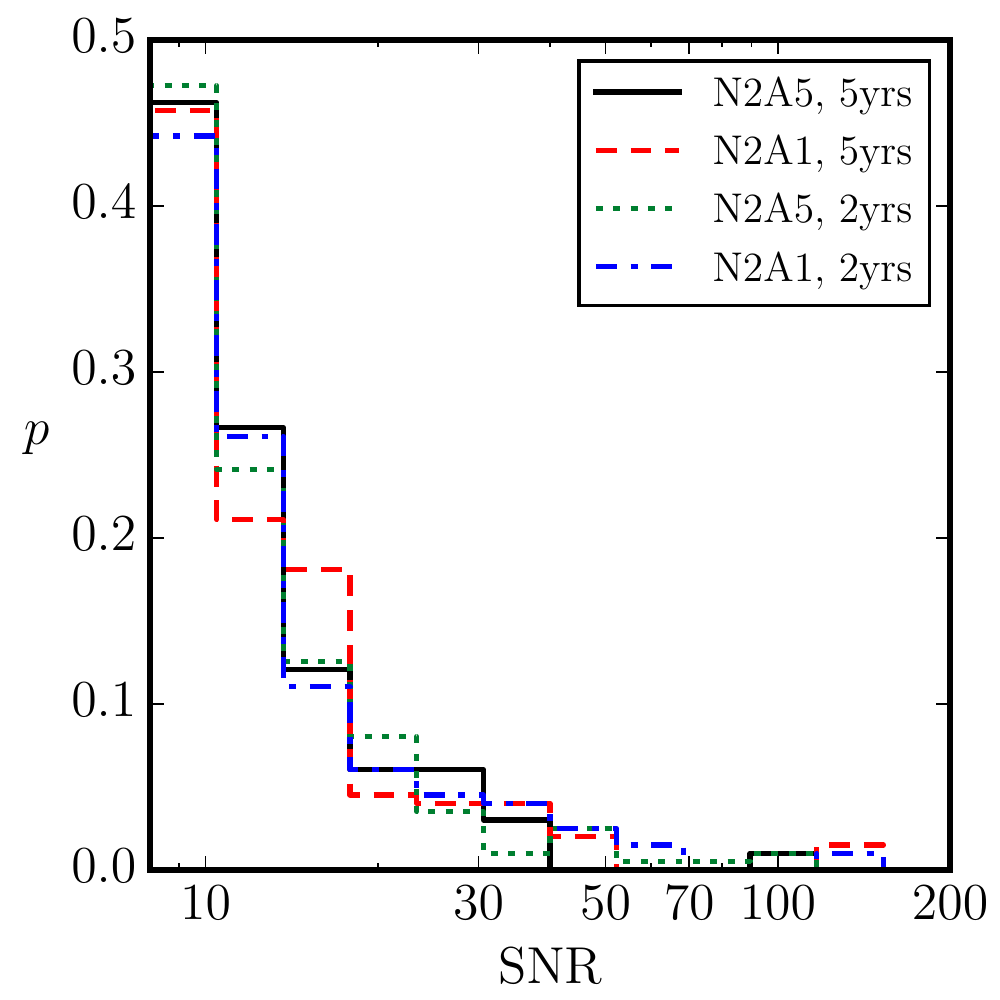}
\caption{Distribution of parameters (from left to right: redshift,
  mass and SNR) for the source catalogs used in our Monte Carlo
  parameter estimation study. The various linestyles correspond to
  N2A5 and $T_{\rm obs}=5\,{\rm yrs}$ (solid black), N2A1 and
  $T_{\rm obs}=5\,{\rm yrs}$ (dashed red), N2A5 and
  $T_{\rm obs}=2\,{\rm yrs}$ (dotted green), N2A1 and
  $T_{\rm obs}=2\,{\rm yrs}$ (dash-dotted blue).}
\label{fig1}
\end{center}
\end{figure*}

\subsection{Plan of the paper}

The rest of the paper provides details on the source catalogs used for
our Monte Carlo simulations, on our waveform models, and on the
parameter estimation errors for other source parameters (including
masses, distance and sky location). In Section~\ref{sec:catalogs} we
describe how we generate the source catalogs used in our Monte Carlo
analysis.  In Section~\ref{sec:wf} we describe our ``restricted'' and
``full'' eccentric waveform models.  In Section~\ref{sec:PE} we show
how eccentricity affects errors on the other parameters (time of
merger, masses, distance and sky location). We conclude with possible
directions for future work. FInally, in Appendix \ref{app:confusion}
we show that confusion noise is unlikely to affect our parameter
estimation calculations. In the whole paper we use geometrical units
($c=G=1$).

\section{Source catalogs}
\label{sec:catalogs}

Following the LIGO/Virgo paper on rate
estimates~\cite{Abbott:2016nhf}, we randomly draw the masses of the
two BHs $m_1$ and $m_2$ from a log-flat mass distribution in the range
$5\,M_{\odot} < m_i < 100\,M_{\odot}$, with the additional requirement
that $M < 100 M_\odot$.  The binary's sky location and the orientation
of the angular momentum are distributed uniformly over the sky. The
source redshift is randomly selected assuming a constant binary BH
merger rate and the Lambda-Cold-Dark-Matter
  ($\Lambda$CDM) flat cosmological model with $\Omega_{\rm m}=0.3$,
$\Omega_{\Lambda}=1-\Omega_{\rm m}$ and
$H_0=
72\,{\rm{km}}\,{\rm{s}}^{-1}\,{\rm{Mpc}}^{-1}$~\cite{Hinshaw:2012fq}.
For each binary we can compute the SNR $\rho$, defined as
\begin{equation}
\rho^2 \equiv 4 \int_{f_{\rm min}}^{f_{\rm max}} \frac{|\tilde{h}(f)|^2}{S_h(f)} df \;,
\end{equation}
where we use analytical approximations to the N2A1 and N2A5 noise
power spectral densities $S_h(f)$ \cite{Klein:2015hvg}.  When
computing SNRs we fix the reference eccentricity $e_0$ to zero:
corrections due to nonzero $e_0$ are of order $e_0^2$, and they are
less than $1\%$ for the fiducial values $e_0 \leq 0.1$ considered in
this paper.

\begin{table}[t]
\begin{center}
\begin{tabular}{ccccccc}
noise & $T_{\rm obs}$& $\bar N$ & 95\% & $\bar z$ & $\bar M$ & $\bar \rho$  \\
\hline
\hline
N2A1 & 2yr &2 &0--8 & 0.0353 & 78.9 & 11.1 \\
N2A1 & 5yr &4 &0--14 & 0.0494 & 80.7 & 11.0 \\
N2A5 & 2yr &30 &5-121  & 0.0803 & 82.8 & 10.6 \\
N2A5 & 5yr &106 &13-348 & 0.170 & 85.0 & 10.9 \\
\hline
\hline 
\end{tabular}
\end{center}
\caption{Bulk properties of the Monte Carlo distributions chosen for
  our study. $\bar N$ denotes the median (and ``95\%''  the 95\%
  confidence interval) of eLISA detections,  given current
  uncertainties in binary BH merger rates. The last three columns list
  median values for the redshift, total mass and SNR.}
\label{tab1}
\end{table}	
 
We generate $N_s=500$ binary BH sources that are observable by eLISA
by imposing a detection threshold $\rho >8$ for each observation
period and noise curve. The mass, redshift and SNR distributions of
the events generated in this way are shown in Fig.~\ref{fig1}, and the
medians of these quantities are listed in Table~\ref{tab1}.  The SNR
and mass distributions are very similar in all four cases, due to the
chosen detection threshold in SNR and to the relatively limited mass
range for the binary components, respectively. With higher detector
sensitivity and longer observation times (corresponding to smaller
$f_{\rm min}$) it is possible to detect sources at higher redshifts,
because the GW amplitude
$\tilde h\sim {\cal M}_z^{5/6} f^{-7/6} D_L^{-1} \sim {\cal M}^{5/6}
f^{-7/6} z^{-1}$
at small redshifts. Note that the tail of the redshift distribution
extends below $z=0.01$, corresponding to $\sim 40\,{\rm Mpc}$, below
which the galaxy distribution is not continuous. The number of sources
we simulated ($N_s=500$) was chosen arbitrarily to study probability
distributions in parameter estimation accuracy. The absolute number of
observed events depends, of course, on binary BH merger rates. In
Table~\ref{tab1} we list the median $\bar N$ and 95\% confidence
interval of expected eLISA detections for each assumed noise curve and
mission duration.

\section{Eccentric binary waveforms}
\label{sec:wf}

The most accurate Fourier-domain eccentric waveforms
  available at present were computed by Yunes et
  al.~\cite{Yunes:2009yz} and Tanay et al.~\cite{Tanay:2016zog} in the
  small-eccentricity approximation, i.e. using a power series
  expansion in $e_0$. The waveforms in~\cite{Yunes:2009yz} are
  accurate up to (Newtonian, $e_0^8$) order in amplitude and
  (Newtonian, $e_0^6$) order in phase. The waveforms
  in~\cite{Tanay:2016zog} used here are accurate up to (Newtonian,
  $e_0^6$) order in amplitude and (2PN, $e_0^6$) order in phase. The
  waveform phase calculation has recently been extended up to 3PN by
  Moore et al.~\cite{Moore:2016qxz}; however their calculation is
  limited to $e_0^0$ order in amplitude and $e_0^2$ order in
  phase. The waveforms in~\cite{Yunes:2009yz,Tanay:2016zog} are more
  accurate for our present purposes, because eLISA observes the
  low-frequency early inspiral of a BH binary, where eccentricity is
  larger (recall that $e\propto f^{-19/18}$) and PN effects are
  relatively less important.

As discussed in the introduction, the sources we are interested in are
expected to have eccentricities $e_0\lesssim 0.1$ at frequencies
$f_0=10^{-2}$~Hz, roughly corresponding to the ``bucket'' of eLISA's
sensitivity window. Therefore we are justified in using the
small-eccentricity waveform generation formalism proposed
in~\cite{Yunes:2009yz} and developed in~\cite{Tanay:2016zog}.
Nonspinning eccentric waveforms depend on ten
  physical parameters
  $\{ {\cal M}_z, \eta, t_c, \phi_c, D_L, e_0, \bar{\theta}_L,
  \bar{\phi}_L, \bar{\theta}_S, \bar{\phi}_S \}$:
  redshifted chirp mass, symmetric mass ratio, time and phase at
  coalescence, luminosity distance, eccentricity at
  $10^{-2}\,{\rm Hz}$, two angles describing the direction of the
  orbital angular momentum, and two angles corresponding to the
  orientation of the source in the sky. The angular variables are
  measured in the solar barycentric frame. This eccentric waveform
is, in general, quite complicated, and for our parameter estimation
calculations we will further expand the frequency-domain waveforms,
first including only phase corrections up to leading order in
eccentricity (what we will refer to as the ``restricted eccentric''
case, Section~\ref{sec:resecc}), and then including up to
next-to-leading order phase corrections as well as amplitude
modulations ("full eccentric'' case, Section~\ref{sec:NLOecc}). As we
will see, restricted eccentric waveforms are useful to gain analytical
understanding of the effects due to nonzero eccentricity, but they are
insufficient for parameter estimation. This happens mainly because
restricted waveforms do not include frequency sidebands to the
dominant harmonic at $f=2f_{\rm orb}$. These sidebands, which are
present in the ``full eccentric'' waveforms, carry crucial information
that is necessary to break parameter degeneracies.

\subsection{Restricted eccentric waveforms}
\label{sec:resecc}

The Fourier transform of the 2PN restricted gravitational waveform for
a nonspinning circular binary with an eccentric-orbit phase
correction reads \cite{Krolak:1995md}
\begin{equation}
\tilde{h} (f) = \frac{A}{D_L(z)} {\cal M}_z^{5/6} f^{-7/6} e^{i \Psi(f)} \left\{ 
\frac{5}{4} {\cal F}_{\alpha} \left[ t(f) \right] \right\} e^{-i \varphi_D\left[ 
t(f) \right]} \;,
\end{equation}
where the amplitude $A= 1/(\sqrt{6}\, \pi^{2/3})$ includes a factor
$\sqrt{3}/2$ because eLISA's arms have an opening angle of $60^\circ$,
as well as a $\sqrt{3/20}$ factor needed to use a sky-averaged
sensitivity~\cite{Berti:2004bd}. Denoting the
  $\alpha$th detector's response functions by $F^{+}_{\alpha}$ and
  $F^{\times}_{\alpha}$, the unit vector of orbital angular momentum
  by $\hat{\mathbf{L}}$, the unit vector directed to the source by
  $\hat{\mathbf{N}}$, and the phase of the detector's orbital motion
  by $\bar{\phi}$, the phasing is given by
\begin{align}
\label{psi:resecc}
\Psi (f) &= 2\pi f \,t_c -\phi_c -\frac{\pi}{4} +\frac{3}{128} 
(\pi {\cal M}_z f)^{-5/3} \nonumber \\
& \times \left[ 1- \frac{2355}{1462} e_0^2 \chi^{-19/9} + \frac{20}{9}
  \left( \frac{743}{336}+\frac{11}{4} \eta \right) x \right. \nonumber \\
& -16\pi x^{3/2} \nonumber \\
&\left. + \left( \frac{15293365}{508032}+ \frac{27145}{504} \eta + 
\frac{3085}{72} \eta^2 \right) x^2 \right] \;, \\
{\cal F}_{\alpha} (t) &= \left\{1+(\hat{\mathbf{L}}\cdot \hat{\mathbf{N}})^2 
\right\} F^{+}_{\alpha}(t)- 2i (\hat{\mathbf{L}}\cdot \hat{\mathbf{N}}) 
F^{\times}_{\alpha}(t)\;, \\
\varphi_{\rm D}(t) &= 2\pi f (t) R \sin \bar{\theta_{\rm S}} \cos [ 
\bar{\phi}(t) - \bar{\phi}_{\rm S} ] \;,
\end{align}
where $R=1\,{\rm AU}$. The time variable $t$ is related to the frequency $f$ by
\begin{align}
t(f) &=t_c -\frac{5}{256} {\cal M}_z (\pi {\cal M}_z f)^{-8/3}
       \nonumber \\
&\times \left[ 1-\frac{157}{43} \frac{e_0^2}{\chi^{19/9}}+\frac{4}{3} \left(
  \frac{743}{336}+\frac{11}{4} \eta \right) x - \frac{32\pi}{5} x^{3/2}\right. \nonumber \\
&\left.
 + \left( \frac{3058673}{508032} + \frac{5429}{504}\eta
  + \frac{617}{72} \eta^2  \right) x^2   \right] \;.
\end{align}

\subsection{Full eccentric waveforms}
\label{sec:NLOecc}

A better approximation to the Fourier transform of the gravitational
waveform for a nonspinning eccentric binary
is~\cite{Yunes:2009yz,Tanay:2016zog}
\begin{align}
\tilde{h} (f) &= \sum_{\ell = 1}^{10}\tilde{h}_\ell (f), \\
\tilde{h}_{\ell} (f) &= \frac{A}{D_L(z)} {\cal M}_z^{5/6} f^{-7/6}
                       e^{i \Psi_{\ell}(f)} \nonumber \\
&\times \left\{ \frac{5}{8} \xi_{\ell} \left[ t(f)
  \right] \left( \frac{\ell}{2} \right)^{2/3} \right\} e^{-i
  \varphi_{{\rm D},\ell}\left[ t(f) \right]} \;,
\end{align}
where
\begin{align}
\xi_{\ell}(t) &= 
\frac{(1-e^2)^{7/4}}{(1+\frac{73}{24}e^2+\frac{37}{96}e^4)^{1/2}} \left\{ 
\Gamma_{\ell}(t)+i \Sigma_{\ell}(t) \right\} \;,\\
\Gamma_{\ell}(t) &= F_{\alpha}^{+}(t) C_{+}^{(\ell)} +F_{\alpha}^{\times}(t) 
C_{\times}^{(\ell)} \;,\\
\Sigma_{\ell}(t) &= F_{\alpha}^{+}(t) S_{+}^{(\ell)} + F_{\alpha}^{\times}(t) 
S_{\times}^{(\ell)} \;,\\
\varphi_{{\rm D},\ell}(t) &= 2\pi \frac{2f}{\ell} R \sin \bar{\theta_{\rm S}} 
\cos [ \bar{\phi}(t) - \bar{\phi}_{\rm S} ] \;.
\end{align}
The coefficients $C_{+}^{(\ell)}$, $C_{\times}^{(\ell)}$,
$S_{+}^{(\ell)}$, $S_{\times}^{(\ell)}$ depend on the eccentricity $e$
and on the inclination angle $\iota$, and they are given
in~\cite{Yunes:2009yz} (where the azimuthal angle determining the
position of the detector relative to the source, $\beta$ in the
notation of~\cite{Martel:1999tm,Yunes:2009yz}, is set to zero).
Here we assume $e_0\ll 1$ and retain terms up to $O(e_0^2)$, with
the following result:
\begin{align}
\tilde{h}_2 (f) &= \frac{A}{D_L(z)} {\cal M}_z^{5/6} f^{-7/6} e^{i \Psi_2(f)} 
\left\{ \frac{5}{4} {\cal F}_{\alpha} \left[ t(f) \right] \right\} e^{-i 
\varphi_{D,2}\left[ t(f) \right]} \;.\\
\tilde{h}_1(f) &= q_1[f,t(f)] e^{i[\Psi_1(f)-\Psi_2(f)]} \tilde{h}_2(f) e^{-i 
\varphi_{D,1}\left[ t(f) \right]}\;, \\
\tilde{h}_3(f) &= q_3(f) e^{i[\Psi_3(f)-\Psi_2(f)]} \tilde{h}_2(f) e^{-i 
\varphi_{D,3}\left[ t(f) \right]} \;,
\end{align}
where
\begin{align}
q_1(t) &= \left( \frac{1}{2} \right)^{8/3} \chi^{-19/18} e_0 \left[ 3- 
\frac{2 \{ 1-(\hat{\mathbf{L}}\cdot \hat{\mathbf{N}})^2 \} F_{\alpha}^{+}(t) {\cal F}_{\alpha}^{\ast}(t)}{|{\cal 
F}_{\alpha}(t)|^2} \right] \;, \\
q_3 &= \left( \frac{3}{2} \right)^{8/3} \chi^{-19/18} e_0 \;.
\end{align}
The 2PN phase up to $O(e_0^2)$ is~\cite{Tanay:2016zog}
\begin{widetext}
\begin{align}
\Psi_{\ell} (f) &= 2\pi f \,t_c - \frac{\ell}{2} \phi_c -\frac{\pi}{4} 
+\frac{3}{128} \left( \frac{\ell}{2} \right)^{8/3}
(\pi {\cal M}_z f)^{-5/3}  \left[ 1- \frac{2355}{1462} e_0^2 \chi^{-19/9} 
\right. \nonumber \\
& + x \left\{  \frac{3715}{756} + \frac{55}{9} \eta  + \left( \left( 
-\frac{2045665}{348096}-\frac{128365}{12432}\eta \right)\chi^{-19/9}+ \left( 
-\frac{2223905}{491232} +\frac{154645}{17544}\eta \right)\chi^{-25/9} 
\right)e_0^2 \right\} \nonumber \\
& + x^{3/2} \left\{ -16\pi + \left( \frac{65561\pi}{4080} \chi^{-19/9} - \frac{295945\pi}{35088} \chi^{-28/9} \right) e_0^2 \right\} \nonumber \\
& + x^2 \left\{ \frac{15293365}{508032}+ \frac{27145}{504} \eta + \frac{3085}{72} \eta^2 + \left( -\frac{111064865}{14141952} -\frac{165068815}{4124736}\eta-\frac{10688155}{294624}\eta^2 \right) \chi^{-19/9} e_0^2 \right. \nonumber \\
& \left. \left.+ \left( -\frac{5795368945}{350880768}+\frac{4917245}{1566432}\eta+\frac{25287905}{447552}\eta^2 \right)\chi^{-25/9}e_0^2 + \left( \frac{936702035}{1485485568}+\frac{3062285}{260064} \eta -\frac{14251675}{631584}\eta^2 \right)\chi^{-31/9}e_0^2 \right\}  \right] \;.
\end{align}
\end{widetext}
The relation between time and frequency up to 2PN can be derived from
Eq.~(B8a) in~\cite{Tanay:2016zog}. Keeping terms up to $O(e_0^2)$, we
can integrate $dF/dt$ and obtain $t(F)$. Setting $F=f/\ell$ with
$\ell=2$, we have
\begin{widetext}
\begin{align}
t(f) &= t_c -\frac{5}{256} \mathcal{M}_z (\pi \mathcal{M}_z f)^{-8/3}
\left[ 1-\frac{157}{43} e_0^2 \chi^{-19/9} \right. \nonumber \\
& + x \left\{  \frac{743}{252} + \frac{11}{3} \eta 
+ \left(\left( -\frac{409133}{37296} - \frac{25673}{1332}\eta \right) 
\chi^{-19/9} + \left( - \frac{444781}{43344} + \frac{30929}{1548} \eta \right) 
\chi^{-25/9} \right) e_0^2 \right\} \nonumber \\
&+ x^{3/2} \left\{ -\frac{32}{5} \pi + \left( \frac{65561 \pi}{2448} 
\chi^{-19/9} - 
\frac{59189 \pi}{3096} \chi^{-28/9} \right) e_0^2
\right\} \nonumber \\
&+ x^2 \left\{ \frac{3058673}{508032} + \frac{5429}{504}\eta
  + \frac{617}{72} \eta^2 + \left( -\frac{22212973}{1928448}  - 
\frac{33013763}{562464} \eta - \frac{2137631}{40176} \eta^2 \right) e_0^2 
\chi^{-19/9}
  \right. \nonumber \\
&+ \left. \left(-\frac{1159073789}{37594368}  + 
\frac{983449}{167832} \eta + \frac{5057581}{47952} \eta^2 \right) e_0^2 
\chi^{-25/9} + \left( \frac{187340407}{131072256} + 
\frac{10411769}{390096} \eta - \frac{2850335}{55728} \eta^2 \right) e_0^2 
\chi^{-31/9}  \right\} \;.
\label{eq:tmerge2PN}
\end{align}
\end{widetext}

\begin{table*}[t]
\begin{center}
\begin{tabular}{ccccccccc}
noise &$T_{\rm obs}$ & $e_0$ & $\Delta \log {\cal M}$ & $\Delta \log \eta$ & 
$\Delta t_c$\,[{\rm s}] & $\Delta \log D_L$ & $\Delta e_0$ & $\Delta 
\Omega_S\,[{\rm deg}^2]$  \\
\hline  
\hline  
N2A1 & 2yr & 0 & $1.49\times 10^{-6}$ & $6.82\times 10^{-3}$ & 1.52 & 
0.438 & --- & $1.06\times 10^{-1}$ \\
&& $10^{-3}$ & $6.97\times 10^{-6}$ & $3.02\times 10^{-2}$ & 2.74 & 0.438 & 
$2.16\times 10^{-3}$ & $1.42\times 10^{-1}$ \\
&& $10^{-2}$ & $6.97\times 10^{-6}$ & $3.02\times 10^{-2}$ & 2.74 & 0.438 & 
$2.16\times 10^{-4}$ & $1.42\times 10^{-1}$ \\
&& $10^{-1}$ & $7.01\times 10^{-6}$ & $3.03\times 10^{-2}$ & 2.75 & 0.437 & 
$2.11\times 10^{-5}$ & $1.42\times 10^{-1}$ \\
\hline
N2A1 & 5yr & 0 & $5.62\times 10^{-7}$ & $2.87\times 10^{-3}$ & 1.75 & 
0.469 & --- & $1.39\times 10^{-1}$ \\
&& $10^{-3}$ & $3.32\times 10^{-6}$ & $1.51\times 10^{-2}$ & 2.25 & 0.469 & 
$6.12\times 10^{-4}$ & $1.60\times 10^{-1}$ \\
&& $10^{-2}$ & $3.32\times 10^{-6}$ & $1.51\times 10^{-2}$ & 2.25 & 0.469 & 
$6.12\times 10^{-5}$  & $1.60\times 10^{-1}$  \\
&& $10^{-1}$ & $3.34\times 10^{-6}$ & $1.54\times 10^{-2}$ & 2.26 & 0.469 & 
$5.81\times 10^{-6}$  & $1.61\times 10^{-1}$  \\
\hline  
N2A5 & 2yr & 0 & $2.14\times 10^{-6}$ & $1.10\times 10^{-2}$ & 2.29 & 
0.473 & --- & $1.29\times 10^{-1}$ \\
&& $10^{-3}$ & $1.05\times 10^{-5}$ & $5.02\times 10^{-2}$ & 5.11 & 0.473 & 
$2.41\times 10^{-3}$ & $1.86\times 10^{-1}$ \\
&& $10^{-2}$ & $1.05\times 10^{-5}$ & $5.02\times 10^{-2}$ & 5.11 & 0.473 & 
$2.41\times 10^{-4}$ & $1.86\times 10^{-1}$ \\
&& $10^{-1}$ & $1.06\times 10^{-5}$ & $5.06\times 10^{-2}$ & 5.13 & 0.473 & 
$2.33\times 10^{-5}$ & $1.86\times 10^{-1}$ \\
\hline   
N2A5 & 5yr	& 0 & $9.01\times 10^{-7}$ & $5.07\times 10^{-3}$ & 3.17 
& 0.529 & --- & $2.32\times 10^{-1}$ \\
&& $10^{-3}$ & $5.80\times 10^{-6}$ & $2.86\times 10^{-2}$ & 4.36 & 0.529 & 
$6.37\times 10^{-4}$ & $3.03\times 10^{-1}$ \\
&& $10^{-2}$ & $5.80\times 10^{-6}$ & $2.87\times 10^{-2}$ & 4.37 & 0.534 & 
$6.36\times 10^{-5}$ & $3.04\times 10^{-1}$  \\
&& $10^{-1}$ & $5.85\times 10^{-6}$ & $2.80\times 10^{-2}$ & 4.37 & 0.530 & 
$5.78\times 10^{-6}$ & $3.05\times 10^{-1}$ \\
\hline 
\hline 
\end{tabular}
\end{center}
\caption{Median parameter estimation errors with restricted eccentric 
waveforms.}
\label{tab:resNS}
\end{table*}

\begin{table*}[t]
\begin{center}
\begin{tabular}{ccccccccc}
noise & $T_{\rm obs}$ & $e_0$ & $\Delta \log {\cal M}$ & $\Delta \log \eta$ & 
$\Delta t_c$\,[{\rm s}] & $\Delta \log D_L$ & $\Delta e_0$ & $\Delta 
\Omega_S\,[{\rm deg}^2]$  \\
\hline  
\hline  
N2A1 & 2yr & 0 & $1.49\times 10^{-6}$ & $6.82\times 10^{-3}$ & 1.52 & 
0.438 & --- & $1.06\times 10^{-1}$ \\
&& $10^{-3}$ & $7.30\times 10^{-6}$ & $3.14\times 10^{-2}$ & 2.71 & 0.436 & 
$2.26\times 10^{-3}$ & $1.19\times 10^{-1}$ \\
&& $10^{-2}$ & $5.00\times 10^{-6}$ & $2.16\times 10^{-2}$ & 2.13 & 0.436 & 
$1.39\times 10^{-4}$ & $1.10\times 10^{-1}$ \\
&& $10^{-1}$ & $1.07\times 10^{-6}$ & $4.60\times 10^{-3}$ & 1.43 & 0.432 & 
$4.52\times 10^{-6}$ & $1.02\times 10^{-1}$ \\
\hline
N2A1 & 5yr & 0 & $5.62\times 10^{-7}$ & $2.87\times 10^{-3}$ & 1.75 & 
0.469 & --- & $1.39\times 10^{-1}$  \\
&& $10^{-3}$ & $3.55\times 10^{-6}$ & $1.57\times 10^{-2}$ & 2.27 & 0.469 & 
$6.26\times 10^{-4}$  & $1.51\times 10^{-1}$ \\
&& $10^{-2}$ & $2.05\times 10^{-6}$ & $8.52\times 10^{-3}$ & 1.94 & 0.464 & 
$3.56\times 10^{-5}$  & $1.41\times 10^{-1}$ \\
&& $10^{-1}$ & $4.10\times 10^{-7}$ & $1.92\times 10^{-3}$ & 1.72 & 0.450 & 
$1.14\times 10^{-6}$  & $1.35\times 10^{-1}$ \\
\hline  
N2A5 & 2yr	 & 0 & $2.14\times 10^{-6}$ & $1.10\times 10^{-2}$ & 2.29 
& 0.473 & --- & $1.29\times 10^{-1}$ \\
&& $10^{-3}$ & $1.09\times 10^{-5}$ & $5.52\times 10^{-2}$ & 5.71 & 0.473 & 
$2.65\times 10^{-3}$ & $1.43\times 10^{-1}$ \\
&& $10^{-2}$ & $5.87\times 10^{-6}$ & $2.70\times 10^{-2}$ & 3.43 & 0.473 & 
$1.16\times 10^{-4}$ & $1.34\times 10^{-1}$ \\
&& $10^{-1}$ & $1.21\times 10^{-6}$ & $6.12\times 10^{-3}$ & 2.23 & 0.463 & 
$3.80\times 10^{-6}$ & $1.26\times 10^{-1}$ \\
\hline   
N2A5 & 5yr	& 0 & $9.01\times 10^{-7}$ & $5.07\times 10^{-3}$ & 3.17 
& 0.529 & --- & $2.32\times 10^{-1}$ \\
&& $10^{-3}$ & $6.29\times 10^{-6}$ & $3.06\times 10^{-2}$ & 4.45 & 0.529 & 
$6.81\times 10^{-4}$ & $2.44\times 10^{-1}$ \\
&& $10^{-2}$ & $2.37\times 10^{-6}$ & $1.14\times 10^{-2}$ & 3.54 & 0.525 & 
$2.47\times 10^{-5}$ & $2.33\times 10^{-1}$ \\
&& $10^{-1}$ & $5.04\times 10^{-7}$ & $2.83\times 10^{-3}$ & 3.14 & 0.505 & 
$9.04\times 10^{-7}$ & $2.25\times 10^{-1}$ \\
\hline 
\hline 
\end{tabular}
\end{center}
\caption{Median parameter estimation errors with full eccentric waveforms.}
\label{tab:fullNS}
\end{table*}

\begin{figure*}[t]
\begin{center}
\includegraphics[width=\textwidth]{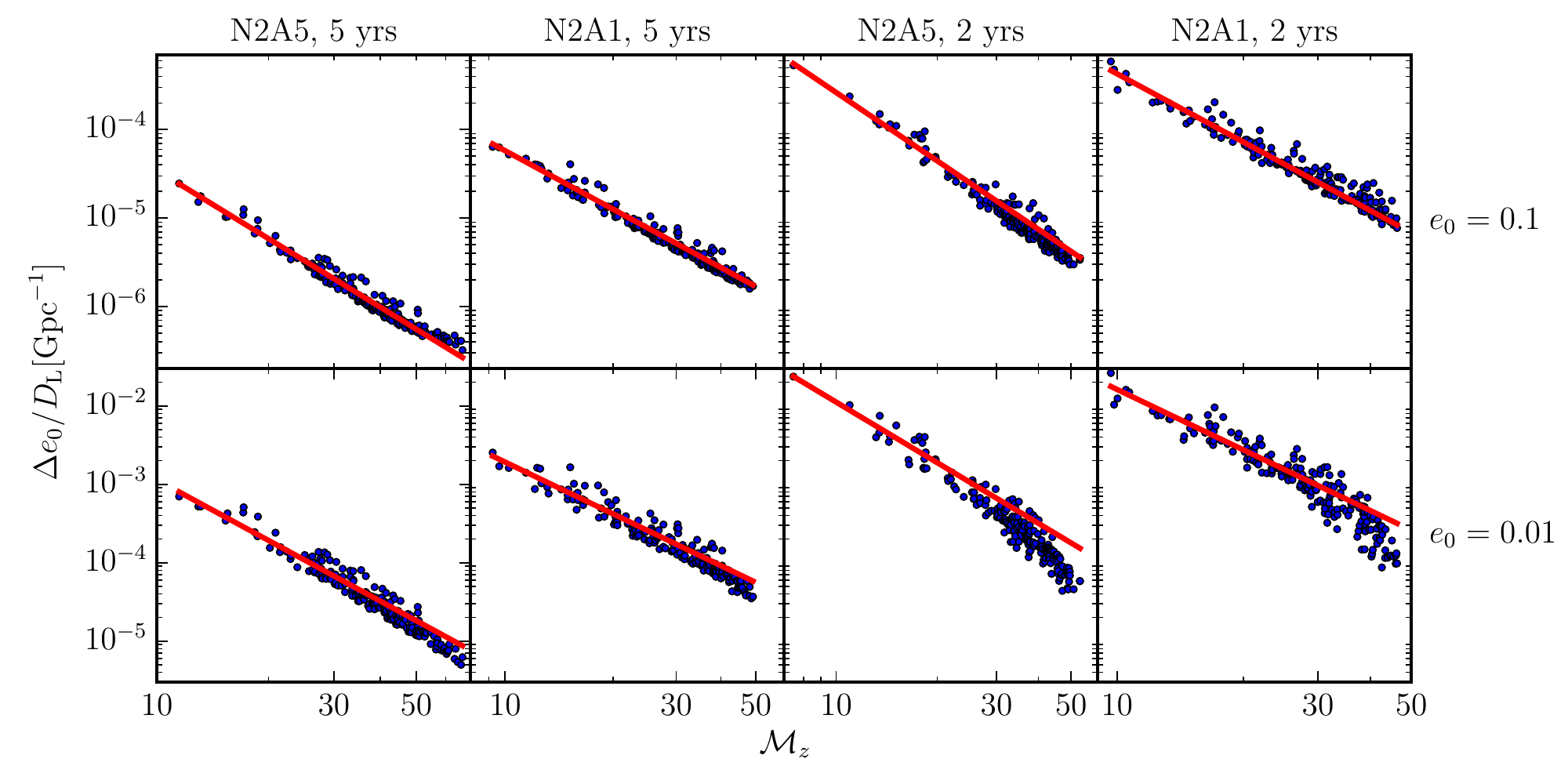}
\caption{Scalings of parameter estimation errors $\Delta e_0/D_L$ with
  ${\cal M}_z$ for full eccentric systems. From top to bottom:
  $e_0 = 0.1$, $e_0 = 0.01$, and from left to right: N2A5, 5 yrs;
  N2A5, 2 yrs; N2A1, 5 yrs; N2A1, 2 yrs. The blue circles represent
  every system in the catalog, and the thick red lines correspond to
  fits for the relation
  $\Delta e_0 / D_{\rm L} = A \mathcal{M}_z^{-\gamma_m}$.}
\label{fig:e0scaling}
\end{center}
\end{figure*}

\begin{figure*}[ht]
\begin{center}
\includegraphics[width=0.49\textwidth]{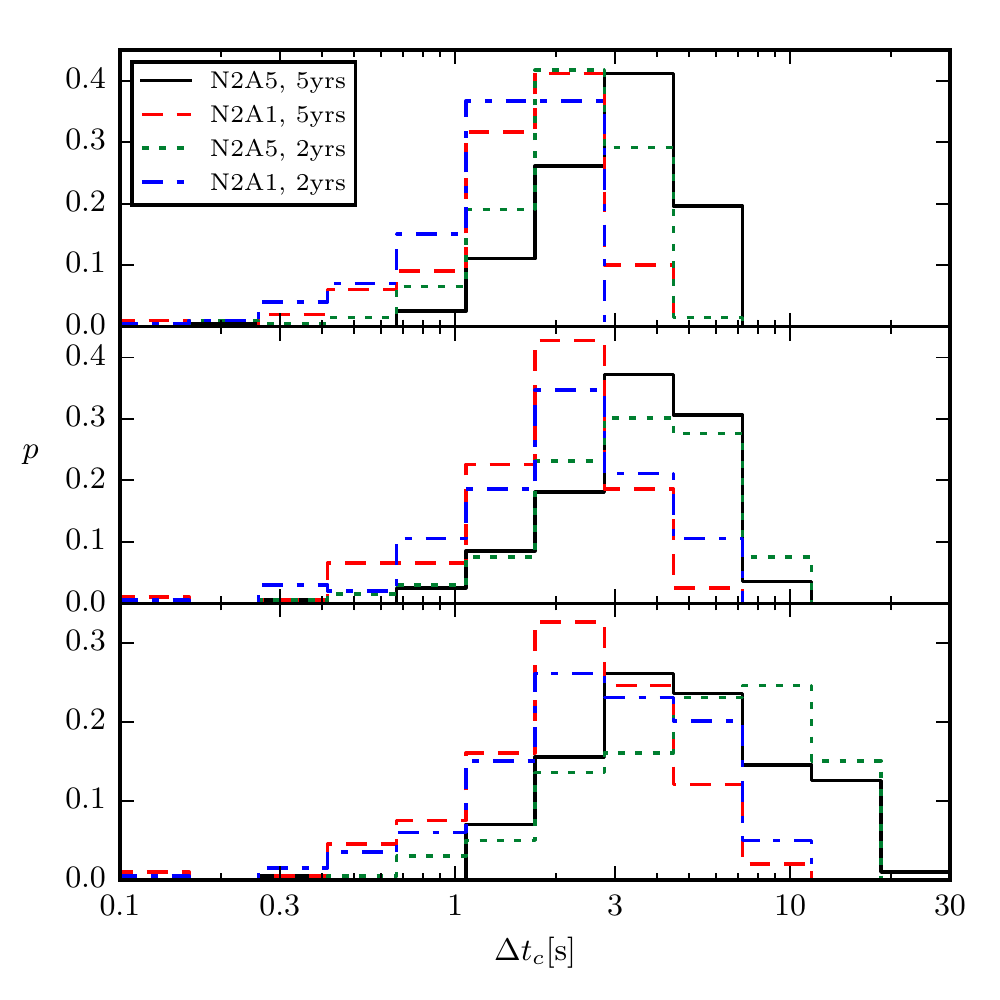}
\includegraphics[width=0.49\textwidth]{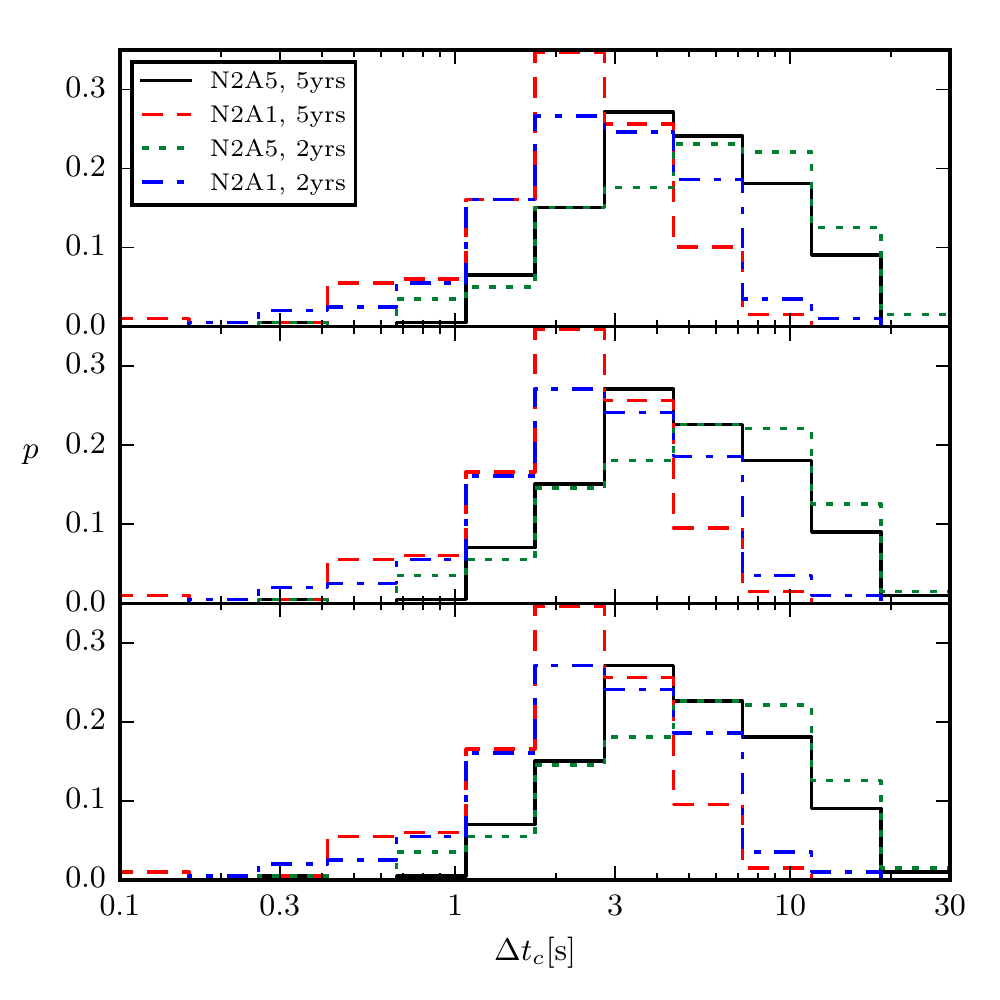}
\caption{Parameter estimation errors on the time of coalescence
  $\Delta t_c$ for full eccentric (left) and restricted eccentric
  (right) nonspinning binaries. On the top, systems with $e_0 = 0.1$,
  in the middle, systems with $e_0 = 0.01$, and at the bottom, systems
  with $e_0 = 0.001$. In solid black, N2A5 with
  $T_{\rm obs}=5\,{\rm yrs}$, in dashed red, N2A1 with
  $T_{\rm obs}=5\,{\rm yrs}$, in dotted green, N2A5 with
  $T_{\rm obs}=2\,{\rm yrs}$, in dash-dotted blue, N2A1 with
  $T_{\rm obs}=2\,{\rm yrs}$.}
\label{fig:tc}
\end{center}
\end{figure*}

\section{Parameter estimation errors}
\label{sec:PE}

Median values of the parameter estimation errors for nonspinning
binaries under different assumptions on the eLISA detector noise and
on the observation time are listed in Table~\ref{tab:resNS} for
restricted eccentric waveforms, and in Table~\ref{tab:fullNS} for full
eccentric waveforms.

Let us focus first on the restricted eccentric parameter estimation
results of Table~\ref{tab:resNS}.
The phasing of the inspiral signal observed by eLISA is predominantly
determined by the mass parameters, which are therefore estimated very
well in most cases. The signal is also modulated by the detector's
orbital motion in a way that depends on the position of the source.
This allows us to determine the sky location of the source and, to
some limited level of accuracy, also the luminosity distance $D_L$
(see e.g.~\cite{Cutler:1997ta,Hughes:2001ya}).
For restricted eccentric waveforms $e_0$ enters only in the phasing
[cf. Eq.~\eqref{psi:resecc}], and therefore it has large correlations
with the mass parameters ${\cal M}$ and $\eta$. As a consequence the
median errors on ${\cal M}$ and $\eta$ are degraded by a factor of 4--6
with respect to the circular case when $e_0\neq 0$.
The estimation errors on the merger time $\Delta t_c$ and sky location
$\Delta \Omega_S$ also get worse by several tens of per cent, but the
degradation in accuracy due to eccentricity is not as large as in the
case of the mass parameters.
Quite remarkably, this degradation in parameter estimation is
independent of $e_0$: the high correlation between the eccentricity
and the mass parameters is not broken by increasing $e_0$ from
$10^{-3}$ to $10^{-1}$.

As shown in Table~\ref{tab:fullNS}, this is not the case for full
eccentric waveforms: the additional structure in the amplitude and
phase due to higher-order effects is crucial to break the
degeneracies.
Once again, a nonzero eccentricity reduces the accuracy in measuring
the other parameters, in particular ${\cal M}$ or $\eta$, whose
determination is degraded by a factor of 4--7 with respect to the
circular case when $e_0=10^{-3}$. However, in stark contrast with the
restricted waveform, as we increase $e_0$ the correlations are
partially broken, and the errors on all parameters (including $e_0$
itself: cf. Fig.~\ref{fig:de0} above) become smaller. In fact, for
$e_0=0.1$ the accuracy in determining the mass parameters becomes
slightly better than in the circular case.
A qualitatively similar (but quantitatively smaller) improvement is
seen in other parameter errors, such as $\Delta t_c$ and
$\Delta \Omega_S$.

Histograms of $\Delta e_0$ for full eccentric waveforms were shown in
the introduction (Fig.~\ref{fig:de0}), where we presented analytical
arguments to justify why $\Delta e_0$ decreases as the chirp mass and
$e_0$ increase. Since frequency sidebands break the correlation
between parameters, parameter estimation errors decrease more rapidly
with $e_0$ in the full eccentric case than in the restricted eccentric
case.
A best fit to our numerical results for $\Delta e_0$ yields the
scaling relation of Eq.~\eqref{scalinge0num}, which is compared
against the data in Fig.~\ref{fig:e0scaling}. The accuracy of the
scaling relation degrades for eLISA designs with shorter armlength and
for shorter mission durations. The scattering of the data is also
larger for small eccentricities, where correlations between $e_0$ and
the other parameters are larger.

\begin{figure*}[t]
\begin{center}
\includegraphics[width=0.49\textwidth]{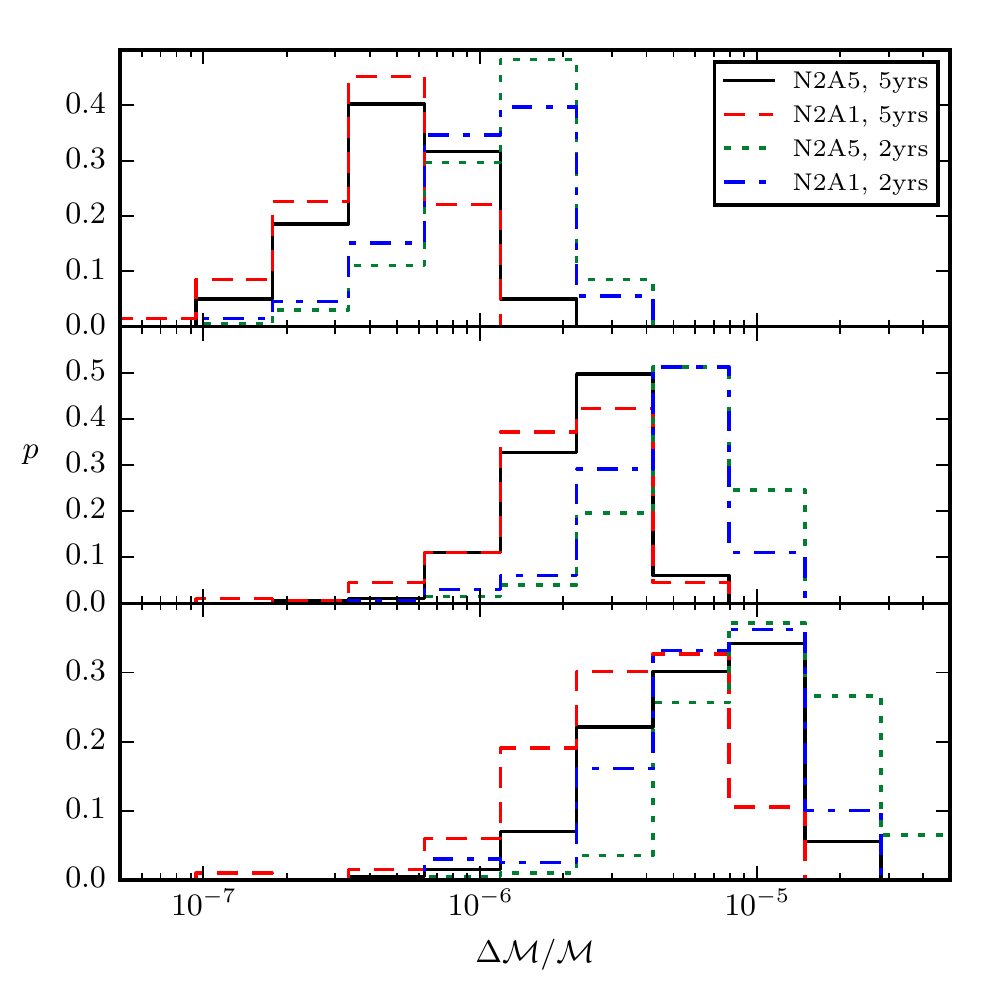}
\includegraphics[width=0.49\textwidth]{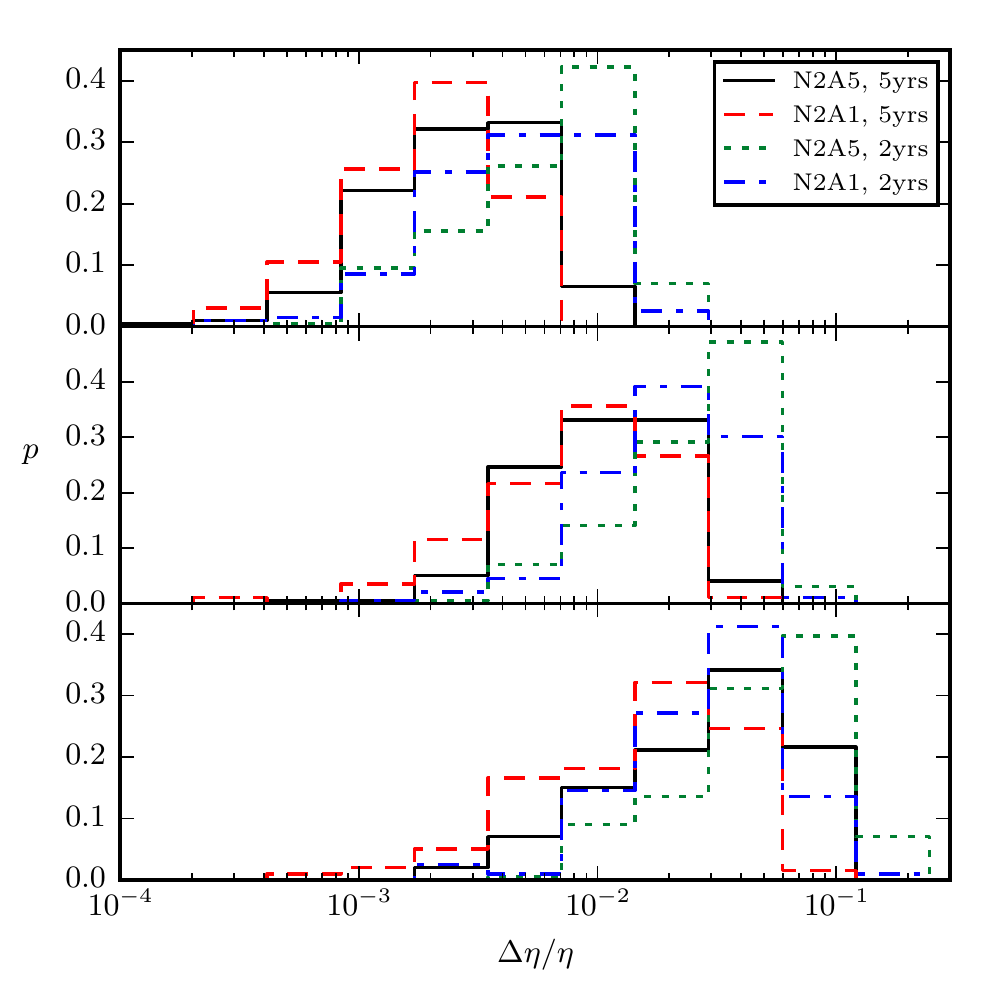}\\
\includegraphics[width=0.49\textwidth]{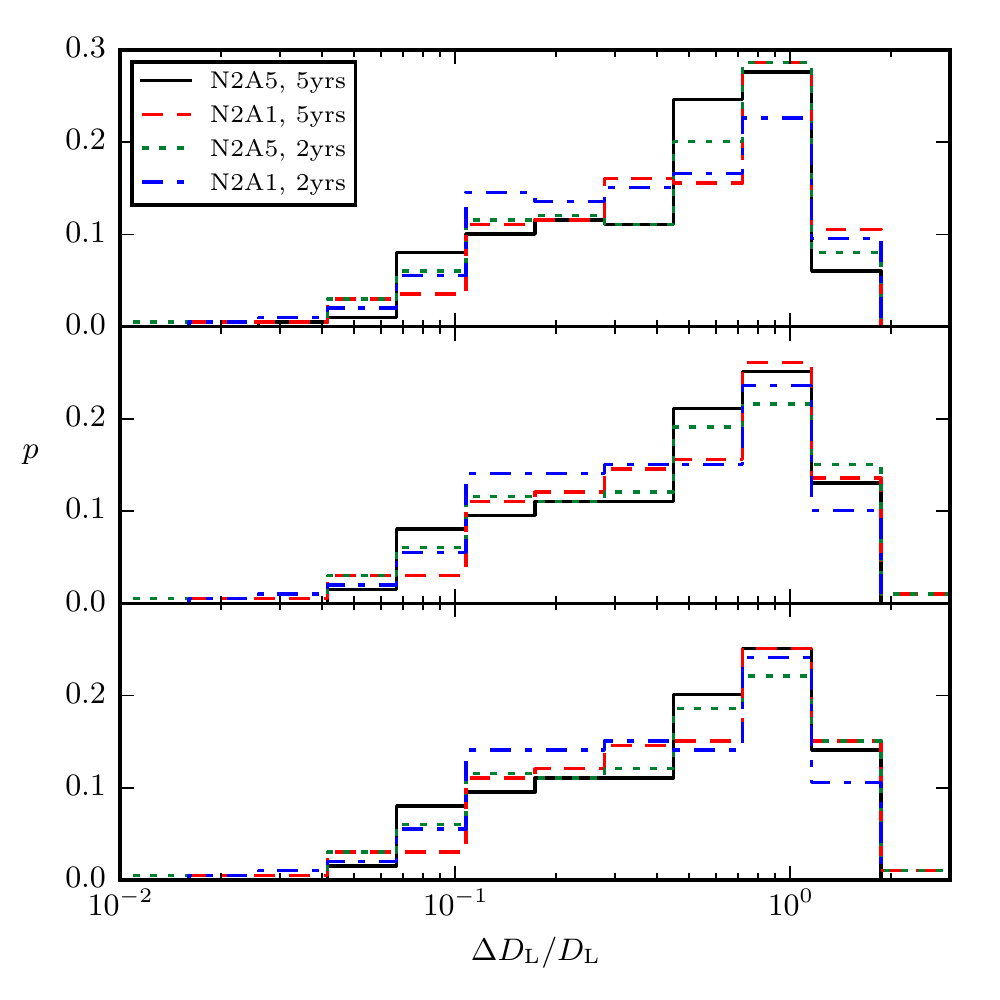}
\includegraphics[width=0.49\textwidth]{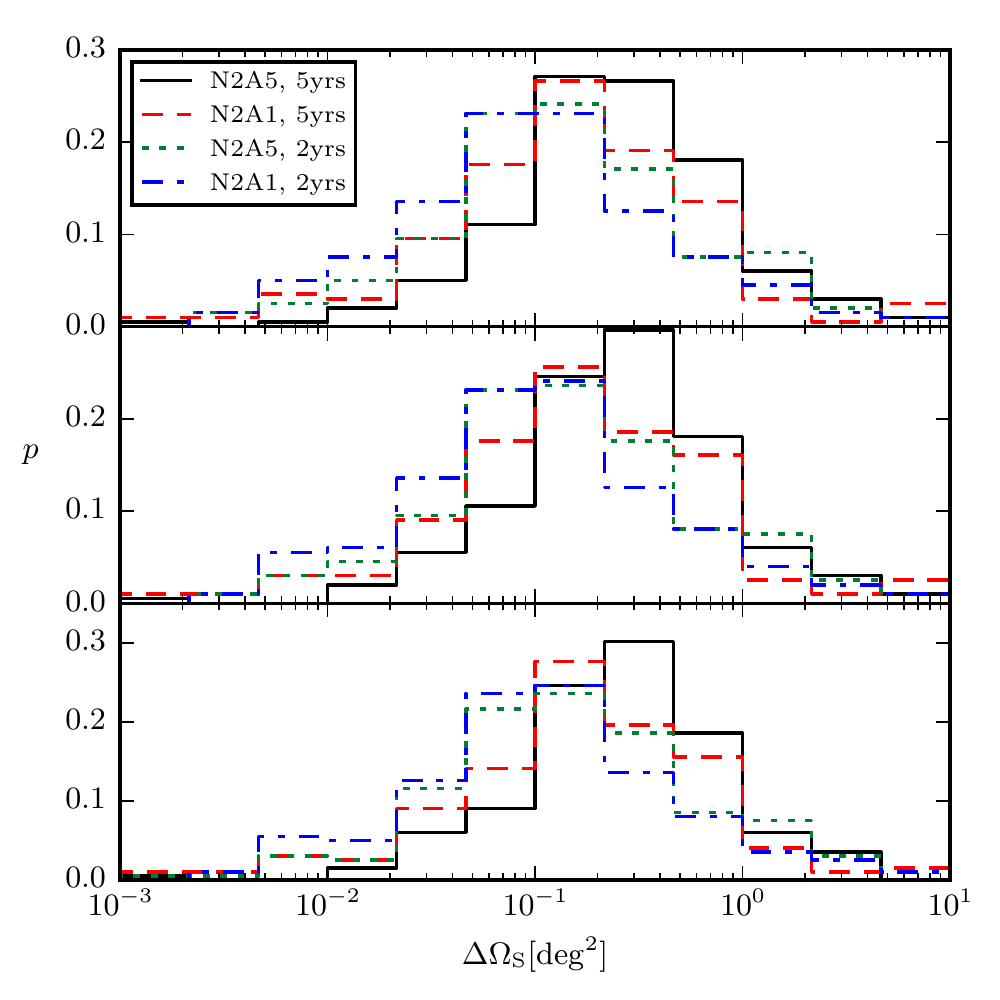}
\caption{Parameter estimation errors on the chirp mass $\mathcal{M}$
  (top left), symmetric mass ratio (top right), luminosity distance
  $D_{\rm L}$ (bottom left) and sky location $\Delta \Omega_{\rm S}$
  (bottom right) for full eccentric nonspinning binaries. On the top,
  systems with $e_0 = 0.1$, in the middle, systems with $e_0 = 0.01$,
  and at the bottom, systems with $e_0 = 0.001$. In solid black, N2A5
  with $T_{\rm obs}=5\,{\rm yrs}$, in dashed red, N2A1 with
  $T_{\rm obs}=5\,{\rm yrs}$, in dotted green, N2A5 with
  $T_{\rm obs}=2\,{\rm yrs}$, in dash-dotted blue, N2A1 with
  $T_{\rm obs}=2\,{\rm yrs}$.}
\label{fig:parsfull}
\end{center}
\end{figure*} 
 
In Fig.~\ref{fig:tc} we compare the error on the merger time for full
(left) and restricted (right) eccentric waveforms. This plot shows
quite clearly that as we increase $e_0$ (bottom to top in each figure)
the determination of $t_c$ gets better in the full eccentric case,
where the more complex waveform breaks the correlation between the
parameters, but not in the restricted eccentric case. This general
trend applies to all measurement errors, so in the following we focus
on full eccentric waveforms.

In Fig.~\ref{fig:parsfull} we use full eccentric waveforms to compute
parameter estimation errors on the chirp mass $\mathcal{M}$ (top
left), symmetric mass ratio $\eta$ (top right), luminosity distance
$D_{\rm L}$ (bottom left) and sky location $\Delta \Omega_{\rm S}$
(bottom right) for full eccentric nonspinning binaries.
The most notable feature of this plot is that the errors on the mass
parameters decrease with $e_0$, while the errors on source localization
and distance are not sensibly affected by $e_0$.

\begin{figure}[h]
\begin{center}
\includegraphics[width=0.49\textwidth]{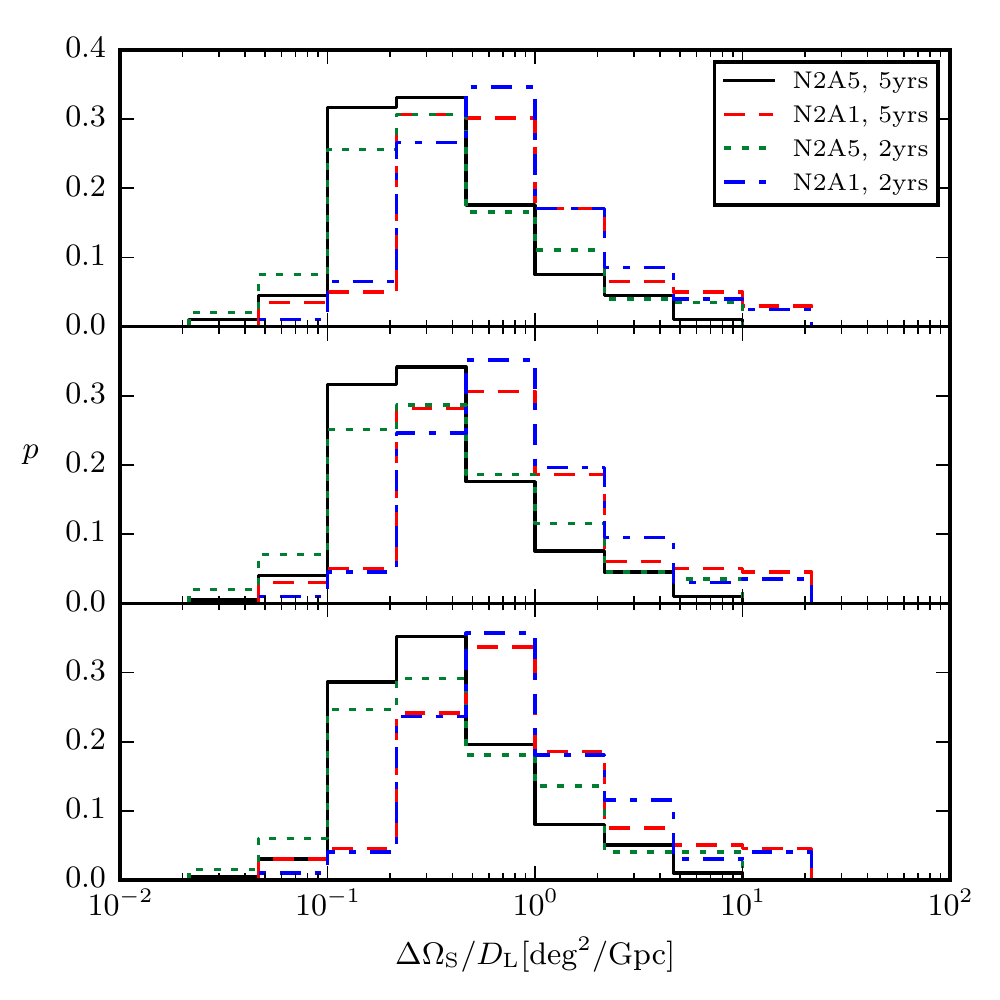}
\caption{Distribution of $\Delta \Omega_S/D_L$ illustrating
  selection effects.}
\label{fig:DOsDL}
\end{center}
\end{figure} 
 
Looking at the sky location determination $\Delta \Omega_S$ in
Fig.~\ref{fig:parsfull}, a careful reader will notice the seemingly
counterintuitive result that binaries observed for 5 years will be
located with worse precision than binaries observed for 2 years. This
is simply a selection effect. Our catalogs were constructed by
imposing an SNR threshold of $\rho>8$, therefore catalogs
corresponding to shorter observation times include systems with
smaller luminosity distance and more optimal orientation. To show that
selection effects are indeed responsible for this counterintuitive
trend, in Fig.~\ref{fig:DOsDL} we plot histograms of the angular
resolution accuracy rescaled by the luminosity distance $D_L$. When
normalized to $D_L$, the angular resolution distributions for the
5-year catalogs are indeed almost indistinguishable from those
computed for the 2-year catalogs.

\section{Discussion}

In this section we discuss how our parameter
  estimation calculations would change if we were to relax some of the
  approximations involved in our waveform models and parameter
  estimation techniques. In particular, we focus on the effect of high
  eccentricity, spins, confusion noise, and the Fisher matrix
  approximation.

\subsection{Highly eccentric binaries}

One important limitation of our approach is the small-$e_0$ expansion
adopted in our waveform models.  All BH binaries we consider are
evolving in frequency above $f=0.01\,{\rm Hz}$, and our results are
accurate at the level of ${\cal{O}}(e_0^2)$. Expected astrophysical
eccentricities for field binaries and binaries in a dense stellar
cluster are $e_0 \lesssim 0.1$. For these populations our phasing is
accurate to within $\sim 1\%$, so we expect our parameter estimation
results to be representative of the capabilities of eLISA when more
accurate waveforms will be available. For binary populations models
which predict large numbers of binaries with $0.1\lesssim e_0<1$,
however, our small-eccentricity approximation is not good enough. In
principle one could keep terms up to ${\cal{O}}(e_0^6)$ using
currently available waveforms, but even the detection of highly
eccentric ($e_0\sim 1$) binaries requires nonperturbative (in $e_0$)
eccentric waveform. The development of accurate high-eccentricity
waveforms is a very active research area and it is beyond the scope of
this
study~\cite{Martel:1999tm,Mikoczi:2012qy,Huerta:2014eca,Sun:2015bva,Forseth:2015oua,Hopper:2015icj,Moore:2016qxz,Loutrel:2016cdw}.

\subsection{Spinning binaries}

In this paper we considered nonspinning BH binaries, but the
introduction of spin parameters in the full eccentric waveforms should
not degrade parameter estimation accuracy. For binaries with aligned
spins, spin effects enter the waveform at 1.5PN order, while
eccentricity enters the waveform at Newtonian level and it is
proportional to $f^{-19/18}$.  This implies that spin effects are more
important at higher frequencies and eccentricity dominates at lower
frequencies, so that degeneracies between spin and eccentricity
effects should be small. In fact we have computed errors on $e_0$ for
nonspinning and aligned-spin binaries using the ``restricted''
eccentric waveforms of Section~\ref{sec:resecc}, and confirmed that
relative variations in the errors are below 60\% (in the worst cases)
for all eLISA configurations considered in this study.

\subsection{Confusion noise}

If many binaries emit in a given observational frequency band, their
signal will constitute a source of confusion noise that can limit
detectability and parameter estimation accuracy. A simple estimate of
this confusion noise is given in Appendix~\ref{app:confusion}, and it
allows us to conclude that our signals are unlikely to be contaminated
by confusion noise.  To verify this statement we can compare the
typical starting frequency of a BH binary for a given eLISA
observation time with the ``confusion noise frequency'' below which
more than two GW signals exist simultaneously in a single frequency
bin. The former is $2.1\times 10^{-2}\,{\rm Hz}$
($1.5\times 10^{-2}\,{\rm Hz}$) for 2-year (5-year) eLISA
observations, respectively. Using Eq.~(\ref{eq34}), the confusion
noise frequency can be estimated to be $1.1\times 10^{-2}\,{\rm Hz}$
($8.4\times 10^{-3}\,{\rm Hz}$) for a typical BH binary merger rate of
$30\,{\rm Gpc}^{-3}\,{\rm yr}^{-1}$ and 2-year (5-year) eLISA
observations, respectively. Therefore, in general, the signal should
be relatively easy to resolve and disentangle in the frequency region
of interest for multiband binaries. In principle extreme mass ratio
inspirals may overlap in frequency with some multiband binaries, but
their waveform is expected to be quite different (because of high
eccentricity and spin precession). Note, moreover, that we do not
expect significant contributions to confusion noise from other
galactic sources (such as WD-WD binaries) at the frequencies of
interest.

\subsection{Fisher matrix analysis}

The Fisher matrix approximation is well known to break down for
low-SNR systems (see e.g.~\cite{Vallisneri:2007ev}). A comparison of
Fisher-matrix results with Markov-Chain Monte Carlo results can be
found in \cite{Rodriguez:2013mla}. Their study focuses on Advanced
LIGO, but their typical SNRs ($\sim 10-20$) are similar to those of
interest in our work. Ref.~\cite{Rodriguez:2013mla} shows that there
Fisher-matrix parameter estimation results have large scatter, but
median values are relatively robust. In this sense, our Fisher
analysis should be relatively reliable for Monte-Carlo studies of
source populations.

In the LISA context, parameter estimation studies beyond the Fisher
matrix approximation were implemented in some studies of WD binaries
end EMRIs, most notably in the Mock LISA Data Challenges
\cite{Babak:2009cj,Blaut:2009si}. Some results of those studies
concern low SNR sources that remain in band for a long time, and they
support the validity of our analysis. The data challenge is to dig out
the signal from the data by matching a sufficient number of cycles,
but once a signal is detected, the precision to which the parameters
are estimated is comparable to Fisher matrix estimates. This has been
demonstrated both for galactic WDs (similar to BH binary signals that
hardly evolve in frequency during the eLISA observation, i.e. those at
frequencies $f<10^{-2}$Hz) and for EMRIs (similar to massive BH
signals chirping and ``crossing over'' to the Advanced LIGO band at
$f>10^{-2}$Hz). In both cases, once the signal is above the detection
threshold (usually assumed to be SNR$=7$ for WD binaries and
SNR$=15-20$ for EMRIs), parameters are estimated with very high
precision and usually also with good accuracy. In a few cases, EMRI
parameters are not accurately recovered because of failures in
identifying the global maximum in the likelihood function, but this is
an issue related to the search algorithms: the likelihood function
exploration fails to correctly identify the sources. This issue is
unlikely to be as relevant here, since BH binary eccentricities are
usually small, implying a smoother behavior of the likelihood function
(prominent secondary maxima associated to strong higher harmonics of
the signal should be absent). In any case, both WD binary and EMRI
parameters have been recovered with high accuracy and precision in the
aforementioned numerical experiments, and the errors are not too far
from Fisher Matrix estimates (usually within a factor of five in the
worst cases).

\section{Conclusions}
\label{sec:concl}

Binaries formed via dynamical interactions in dense star clusters are
expected to be at least mildly eccentric
($e_0\sim 10^{-3}$--$10^{-1}$) at the frequencies $f_0=10^{-2}$~Hz
where eLISA is most sensitive~\cite{Rodriguez:2016kxx}. On the
contrary, binaries formed in the field are expected to have negligible
eccentricities ($e_0\sim 10^{-6}$--$10^{-4}$) in the eLISA
band~\cite{Kowalska:2010qg}. In this paper we carried out Monte Carlo
simulations over a catalog of BH binaries that merge in the Advanced
LIGO band to assess eLISA's potential to measure eccentricity, and
therefore differentiate between competing BH formation scenarios. We
showed that eLISA should always be able to detect a nonzero $e_0$
whenever $e_0\gtrsim 10^{-2}$. If $e_0\sim 10^{-3}$, eLISA will detect
nonzero eccentricity for a fraction $\sim 90\%$ ($\sim 25\%$) of
binaries when the observation time is $T_{\rm obs}=5$ ($2$) years,
respectively. Therefore eLISA observations of BH binaries have the
potential to distinguish between field and cluster formation
scenarios. 

In the future we plan to refine this analysis using better waveform
models and more realistic astrophysical assumptions.  It is
particularly interesting to consider binaries inspiralling at lower
frequencies: these binaries will not necessarily ``cross over'' to the
band accessible by Earth-based detectors, but they may have higher
eccentricity, e.g. because of the Kozai
mechanism~\cite{Wen:2002km,Antonini:2012ad,VanLandingham:2016ccd,Antonini:2015zsa,Samsing:2013kua}. These
highly eccentric systems present a harder challenge in terms of data
analysis, and they motivate further efforts to develop accurate
waveform models and reliable parameter estimation schemes.

\appendix

\section{Confusion noise}
\label{app:confusion}

At low frequencies the frequency evolution of a binary is slower, and
the number of sources in a given frequency bin is larger. If there are
more than two signals simultaneously in a single bin, these signals
are indistinguishable and can produce confusion noise. In this
Appendix we estimate this effect, and we show that confusion noise is
unlikely to affect our conclusions.

The number of inspiral GW signals $\Delta N (f)$ in a bin of frequency
resolution $\Delta f = 1/T_{\rm{obs}}$ is given by
\begin{equation}
\Delta N (f) = \frac{dN}{dt} \left( \frac{df}{dt} \right)^{-1} \Delta f \;. 
\label{eq31} 
\end{equation}  
Here $dN/dt$ is the merger rate per unit time, which can be obtained by integrating over redshift:
\begin{equation}
\frac{dN}{dt} = \int dz \frac{d^2N}{dzdt}= \int \frac{4\pi \chi^2(z)}{(1+z)H(z)} \dot{n}(z) dz \;,
\label{eq32}
\end{equation} 
where $\chi(z)$ is the comoving distance to redshift $z$, and
$\dot{n}(z)$ is the merger rate per unit comoving volume and unit
proper time at redshift $z$. For a constant merger rate
$\dot{n}(z)=\dot{n}_0$, Eq.~(\ref{eq32}) reduces to
\begin{equation}
\frac{dN}{dt} = \dot{n}_0 V \;, \quad \quad V\equiv \int \frac{4\pi \chi^2(z)}{(1+z)H(z)} dz  \;. \nonumber 
\end{equation} 
Substituting the frequency derivative at Newtonian order~\cite{Cutler:1994ys}
\begin{equation}
\frac{df}{dt} = \frac{96}{5} \pi^{8/3} {\cal M}_z^{5/3} f^{11/3} \nonumber
\end{equation}
into Eq.~(\ref{eq31}), we have
\begin{equation}
\Delta N (f) = \frac{5}{96} \pi^{-8/3} \dot{n}_0 V  {\cal M}_z^{-5/3} f^{-11/3} T_{\rm obs}^{-1} \;. \nonumber
\end{equation}
For a power-law mass distribution of the form
\begin{equation}
p({\cal M}) = \frac{{\cal M}^{-\alpha}}{\int_{{\cal M}_{\rm min}}^{{\cal M}_{\rm max}} ({\cal M}^{\prime})^{-\alpha} d{\cal M}^{\prime}} \nonumber
\end{equation}
the number of inspiral GW signals $\Delta N (f)$ should be replaced with the averaged value
\begin{equation}
\langle \Delta N (f) \rangle = \frac{5}{96} \pi^{-8/3} \dot{n}_0 V \langle {\cal M}_z^{-5/3} \rangle f^{-11/3} T_{\rm obs}^{-1} \;, \label{eq33}
\end{equation}
where
\begin{align}
\langle {\cal M}_z^{-5/3} \rangle &= \langle (1+z)^{-5/3} {\cal M}^{-5/3} \rangle \nonumber \\
&= \langle (1+z)^{-5/3} \rangle \int d{\cal M} {\cal M}^{-5/3} p({\cal M}) \nonumber \\
&= \frac{3 (\alpha-1)}{3\alpha+2} \langle (1+z)^{-5/3} \rangle \frac{{\cal M}_{{\rm max}}^{-\alpha-2/3}-{\cal M}_{{\rm min}}^{-\alpha-2/3}}{{\cal M}_{{\rm max}}^{1-\alpha}-{\cal M}_{{\rm min}}^{1-\alpha}}  \nonumber
\end{align}
and (assuming that $\alpha \neq 1$)
\begin{align}
\langle (1+z)^{-5/3} \rangle &= \int_{0}^{\infty} dz \frac{dV/dz}{V} (1+z)^{-5/3} \nonumber \\
&= \frac{1}{V} \int_{0}^{\infty} dz \frac{4\pi \chi^2(z)}{(1+z)^{8/3}H(z)} \;. \nonumber
\end{align}
For a log-flat mass distribution ($\alpha=1$) we would get instead
\begin{equation}
\langle {\cal M}_z^{-5/3} \rangle_{\alpha=1} = \frac{3}{5} \langle (1+z)^{-5/3} \rangle \frac{{\cal M}_{\rm min}^{-5/3}-{\cal M}_{\rm max}^{-5/3}}{\ln [{\cal M}_{\rm max}/{\cal M}_{\rm min}]} \;. \nonumber
\end{equation}
Setting $\langle \Delta N (f_{\rm{conf}}) \rangle=1$ in
Eq.~(\ref{eq33}), we obtain the critical frequency below which more
than two signals are in the same frequency bin:
\begin{equation}
f_{\rm{conf}} = \left( \frac{5}{96} \frac{\dot{n}_0 V}{T_{\rm obs}} \right)^{3/11} \pi^{-8/11} \langle {\cal M}_z^{-5/3} \rangle^{3/11} \;.
\end{equation}
This is the main result of this appendix.  For a $\Lambda$CDM
cosmology with $\Omega_{\rm m}=0.3$,
$\Omega_{\Lambda}=1-\Omega_{\rm m}$, and
$H_0=
72\,{\rm{km}}\,{\rm{s}}^{-1}\,{\rm{Mpc}}^{-1}$~\cite{Hinshaw:2012fq}
and for the astrophysical population considered in this paper, the
averaged quantities are $\langle (1+z)^{-5/3} \rangle \approx 0.121$
and
$\langle {\cal M}_z^{-5/3} \rangle^{-3/5}_{\alpha=1} \approx
35\,M_{\odot}$
(where for simplicity we set $\eta=1/4$). Then the confusion noise
frequency is
\begin{align}
f_{\rm{conf}} &\approx 8.4 \times 10^{-3} \left( \frac{5\,{\rm yr}}{T_{\rm obs}} \right)^{3/11} \left( \frac{\dot{n}_0}{30\,{\rm Gpc}^{-3}\,{\rm yr}^{-1}} \right)^{3/11} \nonumber \\
&\quad \times \left( \frac{35 \,M_{\odot}}{\langle {\cal M}_z^{-5/3} \rangle^{-3/5}} \right)^{5/11} \, {\rm Hz} \;. \label{eq34}
\end{align}

It is useful to estimate the confusion noise power spectrum: even if
there is confusion noise in a given frequency band, its effects can be
ignored as long as the confusion noise amplitude is much smaller than
the detector strain sensitivity.  The energy density of GWs per
logarithmic frequency bin normalized by the critical energy density of
the Universe at the present time can be written
as~\cite{Phinney:2001di}
\begin{align}
\Omega_{\rm gw} (f) &= \frac{8 \pi^{5/3}}{9 H_0^2} {\cal M}^{5/3} f^{2/3} \int_0^{\infty} \frac{\dot{n}(z)}{(1+z)^{4/3} H(z)} dz \;. \nonumber
\end{align}
Using the relation between $\Omega_{\rm gw}$ and the power spectral density \cite{Maggiore:1999vm}
\begin{equation}
\Omega_{\rm{gw}}(f) = \frac{4\pi^2 f^3}{3H_0^2} S_{\rm{h}}(f) \;, \nonumber
\end{equation}
we get the confusion noise power spectral density
\begin{equation}
S_h^{\rm conf} (f) =  \frac{2}{3\pi^{1/3}} {\cal M}^{5/3} f^{-7/3} \int_0^{\infty} \frac{\dot{n}(z)}{(1+z)^{4/3} H(z)} dz \;. \nonumber
\end{equation}
For a power-law mass distribution, ${\cal M}$ should be replaced with
the averaged value
\begin{align}
\langle {\cal M}^{5/3} \rangle &= \int d{\cal M} {\cal M}^{5/3} p({\cal M}) \nonumber \\
&= \frac{3(1-\alpha)}{8-3\alpha} \frac{{\cal M}_{\rm max}^{8/3-\alpha}-{\cal M}_{\rm min}^{8/3-\alpha}}{{\cal M}_{\rm max}^{1-\alpha}-{\cal M}_{\rm min}^{1-\alpha}} \;, \nonumber
\end{align}
where we assumed $\alpha \neq 1$. Using a log-flat mass distribution
($\alpha=1$) yields instead
\begin{equation}
\langle {\cal M}^{5/3} \rangle_{\alpha=1} = \frac{3}{5} \frac{{\cal M}_{\rm max}^{5/3}-{\cal M}_{\rm min}^{5/3}}{\ln [{\cal M}_{\rm max}/{\cal M}_{\rm min}]} \;. \nonumber
\end{equation}
For the astrophysical populations we consider, the averaged quantities
are
$\langle {\cal M}^{5/3} \rangle^{3/5}_{\alpha=1} \approx
19\,M_{\odot}$
and
$\langle {\cal M}^{5/3} \rangle^{3/5}_{\alpha=2.35} \approx
11\,M_{\odot}$
(once again, for simplicity, we set $\eta=1/4$). Assuming a constant
merger rate $\dot{n}(z)=\dot{n}_0$, the confusion noise power spectrum
is
\begin{align}
S_h^{\rm conf} (f) &= 1.2 \times 10^{-42} \left( \frac{10^{-2}\,{\rm Hz}}{f} \right)^{7/3} \left( \frac{\dot{n}_0}{30\,{\rm Gpc}^{-3}\,{\rm yr}^{-1}} \right) \nonumber \\
&\quad \times \left( \frac{\langle {\cal M}^{5/3} \rangle^{3/5}}{19 \,M_{\odot}} \right) \; {\rm Hz}^{-1} \;,
\end{align}
or
\begin{align}
\Omega_{\rm{gw}}(f) &=2.8 \times 10^{-12} \left( \frac{f}{10^{-2}\,{\rm Hz}} \right)^{2/3} \left( \frac{\dot{n}_0}{30\,{\rm Gpc}^{-3}\,{\rm yr}^{-1}} \right) \nonumber \\ 
& \quad \times \left( \frac{\langle {\cal M}^{5/3} \rangle^{3/5}}{19 \,M_{\odot}} \right) \;.
\end{align}
Using the typical parameters considered in this paper, we conclude
that the confusion noise is smaller than the eLISA noise power
spectral density at frequencies above $10^{-2}\,{\rm Hz}$. At
frequencies lower than $10^{-2}\,{\rm Hz}$ some contamination from
confusion noise is possible, depending on eLISA design choices.

\begin{acknowledgments}
  E.B., A.K. and A.N. are supported by NSF CAREER Grant No.
  PHY-1055103 and by NSF Grant No. PHY-1607130.  E.B. and A.K. are
  supported by FCT contract IF/00797/2014/CP1214/CT0012 under the
  IF2014 Programme.  A.S. is supported by a University Research
  Fellowship of the Royal Society. This work was supported by the
  H2020-MSCA-RISE-2015 Grant No. StronGrHEP-690904.
\end{acknowledgments}


%

\end{document}